\documentclass[prx,twocolumn,notitlepage,superscriptaddress,longbibliography]{revtex4-2}
\pdfoutput=1
			
\usepackage{graphicx}			
\usepackage{amsmath}			
\usepackage{amsfonts}
\usepackage{amssymb}
\usepackage{amsthm}
\usepackage{subfigure}

\newcommand{\eq}[1]{Eq.~\eqref{#1}}

\newcommand{\dslash}{\not{\hbox{\kern-2pt $\partial$}}}
\newcommand{\bqa}{\begin{eqnarray}} 
\newcommand{\eqa}{\end{eqnarray}}
\newcommand{\nn}{\nonumber \\}

\newcommand{\beq}{\begin{equation}}
\newcommand{\eeq}{\end{equation}}

\renewcommand{\vec}[1]{{\mathbf{#1}}}

\def\be{\begin{eqnarray}}
\def\ee{\end{eqnarray}}

\def\part{\partial}

\def\Dslash{\,\,{\raise.15ex\hbox{/}\mkern-13mu D}}
\def\Dbarslash{\,\,{\raise.15ex\hbox{/}\mkern-12mu {\bar D}}}
\def\delslash{\,\,{\raise.15ex\hbox{/}\mkern-10mu \partial}}
\def\delbarslash{\,\,{\raise.15ex\hbox{/}\mkern-9mu {\bar\partial}}}
\def\pslash{\,\,{\raise.15ex\hbox{/}\mkern-11mu p}}
\def\qslash{\,\,{\raise.15ex\hbox{/}\mkern-9mu q}}
\def\kslash{\,\,{\raise.15ex\hbox{/}\mkern-11mu k}}
\def\eslash{\,\,{\raise.15ex\hbox{/}\mkern-9mu \epsilon}}

\newcommand{\slsh}[1]{\,\,{\raise.15ex\hbox{/}\mkern-12mu {#1}}}

\usepackage[papersize={8.5in,11in}]{geometry}

\usepackage{color}
\definecolor{darkblue}{rgb}{0.,0.,0.4}
\definecolor{darkred}{rgb}{0.5,0.,0.}
\definecolor{BlueViolet}{RGB}{138,43,226}
\definecolor{SkyBlue}{RGB}{30,144,255}
\definecolor{DarkGreen}{RGB}{0,100,0}
\usepackage[pdftex,colorlinks=true,linkcolor=darkblue,citecolor=blue,urlcolor=darkred]{hyperref}

\begin{document}

\title{Ultraviolet/infrared mixing in non-Fermi liquids}

\author{Ipsita Mandal} 
\affiliation{Perimeter Institute for Theoretical Physics, 31 Caroline St. N., Waterloo ON N2L 2Y5, Canada}

\author{Sung-Sik Lee}
\affiliation{Perimeter Institute for Theoretical Physics, 31 Caroline St. N., Waterloo ON N2L 2Y5, Canada}
\affiliation{Department of Physics $\&$ Astronomy, McMaster University,
1280 Main St. W., Hamilton ON L8S 4M1, Canada}

\begin{abstract}

We study low-energy effective field theories for non-Fermi liquids 
with Fermi surfaces of general dimensions and co-dimensions.
When the dimension of Fermi surface is greater than one,
low-energy particle-hole excitations remain strongly coupled with each other across the entire Fermi surface. In this case, even the observables 
that are local in the momentum space (such as the Green's functions) 
become dependent on the size of the Fermi surface in singular ways, 
resulting in an ultraviolet/infrared (UV/IR) mixing.
By tuning the dimension and co-dimension of the Fermi surface independently,
we find perturbative non-Fermi liquid fixed points
controlled by both UV/IR mixing and interactions.  

\end{abstract}

\maketitle

\section{Introduction}
There have been intensive efforts 
to understand unconventional metallic states 
that lie outside the framework of Laudau Fermi liquid theory\cite{
holstein,
reizer,
leenag,
HALPERIN,
polchinski,
ALTSHULER,
YBKim,
nayak,
lawler1,
SSLee,
metlsach1,
metlsach,
chubukov1,
Chubukov,
mross,
Jiang,
Shouvik1,
Lee-Dalid,
shouvik2}.
Among the goals are to construct minimal field theories
that capture universal low-energy physics,
understand the dynamics in controlled ways,
and eventually come up with a systematic classification 
for non-Fermi liquids.

Non-Fermi liquids can arise when a gapless boson is coupled with a Fermi surface.
One of the important criteria that determines the universal properties of non-Fermi liquids
is the momentum carried by the critical boson.  
If the critical boson carries zero momentum,
fermions lose coherence across the entire Fermi surface. 
The examples include the Ising-nematic critical point \cite{metlsach1,ogankivfr,metzner,delanna,kee,lawler1,rech,wolfle,maslov,quintanilla,yamase1,yamase2,halboth,
jakub,zacharias,kim,huh} and the Fermi surface coupled with an emergent gauge field \cite{MOTRUNICH,LEE_U1,PALEE,MotrunichFisher}. When the critical boson carries a finite momentum at the spin density wave (SDW) or charge density wave (CDW) critical points \cite{metlsach,chubukov1,Chubukov,shouvik2}, 
the electrons on hot spots (or hot lines) play a special role 
because they remain strongly coupled with the critical boson in the low-energy limit. 
Another important criterion that characterizes different types of non-Fermi liquids
is the geometry of Fermi surface.
Although non-Fermi liquids do not have a finite jump in the electron occupation number, 
Fermi surface can be still well-defined through weaker non-analyticities (such as power-law singularities) of the electron spectral function \cite{senthil}.
The Fermi surface, identified from a non-analyticity of the spectral function in a non-Fermi liquid state, 
is inherited from that of the underlying Fermi liquid before the coupling with a gapless boson is turned on.
The kinematic constraints imposed by the parent Fermi surface geometry are 
important in determining the nature of the resulting non-Fermi liquid.
In this paper, our goal is to understand how the nature of non-Fermi liquids depends on the 
Fermi surface geometry  for those cases where the critical boson carries zero momentum.

When the shape of the Fermi surface is globally convex in momentum space,
there is no special point on the Fermi surface \cite{Shouvik1}.
Then Fermi surface geometry is classified 
in terms of the dimension and co-dimension of Fermi surface.
Throughout the paper, we will use $m$ for the dimension of 
Fermi surface and $d$ for the space dimension. 
The co-dimension of Fermi surface is then $d-m$.
$d$ controls the strength of quantum fluctuations,
and $m$ controls the extensiveness of gapless modes.
Although $d$ and $m$ are discrete in reality,
we will treat them as continuously tunable parameters
to find controlled examples
from which physical dimensions can be approached.
Regarding $d$, theories below upper critical dimensions
flow to interacting non-Fermi liquids at low energies,
whereas systems above upper critical dimensions
are expected to be described by Fermi liquids.
Concerning non-Fermi liquids below upper critical dimensions,
theories with $m = 1$ are fundamentally different 
from those with $m>1$.
This is due to an emergent 
locality in momentum space that is present for $m=1$, 
but not for $m>1$ \cite{LEE2008}.
The locality has to do with the fact that
observables that are local in momentum space,
such as Green's functions, can be extracted
from local patches in momentum space
without having to refer 
to global properties of Fermi surface \footnote{If there is a superconducting instability,
the locality in momentum space breaks down even for $m=1$. Here we assume $d-m>1$, for which there is no perturbative pairing instability.}.
By exploiting the locality in momentum space, 
controlled non-Fermi liquid fixed points are found 
in patch descriptions for $m=1$ \cite{nayak,mross,Lee-Dalid,Shouvik1,shouvik2}.

In contrast to the case with $m=1$, 
non-Fermi liquids with $m>1$ are less well-understood.
The naive scaling based on the patch description breaks down 
as the size of Fermi surface ($k_F$) qualitatively
modifies the scaling through the Landau damping. 
This is due to a UV/IR mixing, where
low-energy physics is affected 
by gapless modes on the entire Fermi surface 
in a way that their effects cannot be incorporated
within the patch description
through renormalization of local properties
of the Fermi surface.
In this sense, $k_F$ becomes a `naked scale' for $m>1$. 

Let us elaborate on the origin of the UV/IR mixing.
In a renormalizable relativistic quantum field theory, 
the UV cut-off enters the low-energy
effective theory only through the renormalized parameters
that can be defined at a momentum scale far below the UV cut-off. 
For example, in QED, 
one can extract any observable at a momentum $k_1 << \Lambda$ solely
in terms of  the renormalized mass and charge measured at another momentum  $k_2 << \Lambda$
to the leading order in $k_1/\Lambda$, 
where $\Lambda$ is a large momentum cut-off.
This insensitivity of the long-distance physics to the short-distance physics  
does not necessarily hold in the presence of Fermi surface.
When $m>1$, low-energy observables defined at a momentum near Fermi surface (such as Green's functions)
cannot be described solely in terms of the effective couplings defined near that momentum.

When a critical boson, 
that is coupled to Fermi surface,
has a large number of flavors 
or velocity much higher than the Fermi velocity,
one can find a wide range of energy scale
over which the effect of $k_F$ can be ignored\cite{Fit}.
However, the UV/IR mixing becomes eventually important, 
in the low-energy limit as long as the the number of flavors
and the velocity ratio are finite,
either through the Landau damping 
or through a superconducting instability \cite{Torroba}.
In this paper, we provide a controlled analysis 
that shows how interactions and the UV/IR mixing interplay
to determine  low-energy scalings in non-Fermi liquids with general $m > 1$.

The paper is organized as follows: In Sec.~\ref{model}, we introduce a theory which describes the Ising-nematic quantum critical point for a system with an $m$-dimensional Fermi surface 
embedded in $d$ spatial dimensions.
For general $m$, we identify the upper critical dimension $d_c (m)$ 
at which one-loop quantum corrections exhibit logarithmic divergences. 
Using $\epsilon = d_c(m)-d$ as  an expansion parameter,
one can perturbatively access the non-Fermi liquid states 
that arise in $d< d_c(m)$. 
Sec.~\ref{dr} is devoted to the description of the dimensional regularization scheme 
and the beta function that describes the flow to the non-Fermi liquid fixed point.
In Sec.~\ref{rg}, the RG equations for the renormalized Green's functions are derived.
Based on the one-loop results, the dynamical critical exponent, anomalous dimensions and two-point functions are also computed. 
Sec.~\ref{dimexp} discusses the physical manifestations of UV/IR mixing in physical dimensions.
In Sec.~\ref{ep}, we demonstrate that the expansion is controlled
in the small $\epsilon$ limit with fixed $N$, where $N$ is the number of fermion flavors. 
In particular, we show that the one-loop critical exponents
are not modified by the two-loop (and possibly by all higher-loop) corrections for $m>1$
due to the UV/IR mixing.
We finish with a summary and some outlook in Sec.~\ref{conclusion}. 
Details on the computation of the Feynman diagrams upto two-loop order can be found in the appendix.

\begin{figure}[ht]
\begin{center}
\subfigure[]{\includegraphics[scale=0.6]{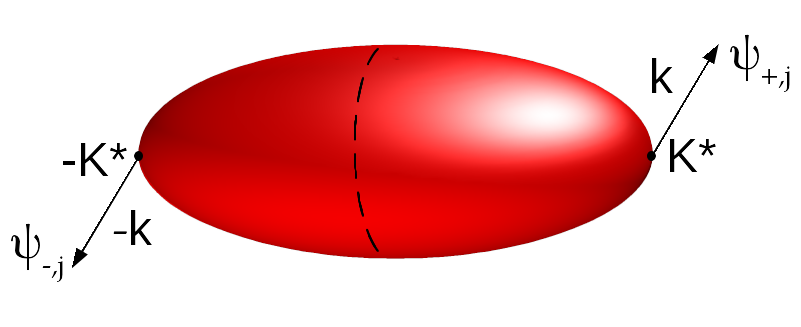} 
\label{fig:FS}} \\
\subfigure[]{\includegraphics[scale=0.6]{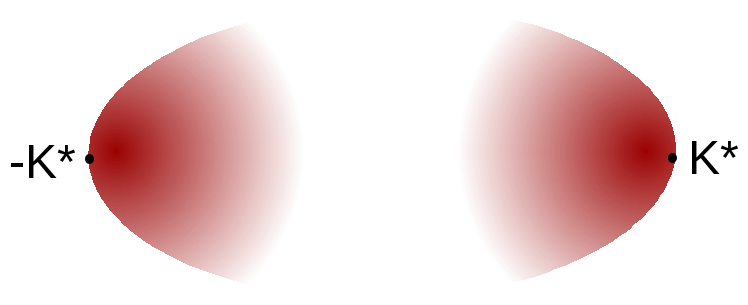} 
\label{fig:patches}} \\
\end{center}
\caption{
(a) A compact Fermi surface can be divided into two halves 
centered at $\pm K^*$. 
For each half, a separate fermionic field is introduced.
(b) The compact Fermi surface is approximated by 
two sheets of non-compact Fermi surfaces 
with a momentum regularization
that suppresses modes far away from $\pm K^*$.
}
\end{figure}


\section{Model}
\label{model}
We first consider an $m$-dimensional Fermi surface, 
which is coupled with a critical boson 
whose momentum is centered at $Q=0$ 
in $d=(m+1)$ space dimensions.
One way of characterizing non-Fermi liquids is
through scaling behaviors of the fermion and boson Green's functions.
For this purpose, it is convenient to focus on the point (say $K^*$) 
at which the fermion Green's function is defined.
At low energies, fermions are mainly scattered 
along the tangential directions by the critical boson.
In the presence of the inversion symmetry,
fermions near $K^*$ are most strongly
coupled with fermions near its anti-podal point $-K^*$,
whose tangent space coincides with that of $K^*$.
With this in mind, we divide the closed Fermi surface into two halves centred at momenta $K^*$  and $-K^*$ respectively, 
and introduce separate fermionic fields $\psi_{+,j}$ and $\psi_{-,j}$ representing the corresponding halves, as shown in Fig. \ref{fig:FS}.
In this coordinate system, the action is written as
\bqa
S & =&   \sum_{s=\pm,j}  \int dk \,
\psi_{s,j}^\dagger (k)
\Bigl[ 
 i k_0   +  s  k_{1} +   {\vec L}_{(k)}^2 + H( {\vec L}_{(k)}^2 )  \Bigr] \psi_{s,j}(k) \nonumber \\
 &+& \frac{1}{2} \int  dk
 \left[ k_0^2 + k_{1}^2 +  {\vec L}_{(k)}^2 \right] \phi(-k) \, \phi(k) \nonumber \\
 &+&  \frac{1}{\sqrt{N}} \sum_{s=\pm,j}   
\int dk \, dq ~  e_s \phi(q) ~  \psi^\dagger_{s,j}(k+q) \, \psi_{s,j}(k) \, .
\label{act0-2}
\eqa
Here $k$ is the $(d+1)$-dimensional energy-momentum vector with
$dk \equiv \frac{d^{d+1} k}{(2\pi)^{d+1} }$. 
$\psi_{+,j}(k_0,k_i)$ $\left ( \psi_{-,j}(k_0,k_i) \right)$ represents the fermion field
with flavor $j=1,2,..,N$, frequency $k_0$ and momentum $K_i^*+k_i$ ($-K_i^*+k_i$) with $1 \leq i \leq d$.
$k_{1}$ and  ${\vec L}_{(k)} ~\equiv~ (k_{2}, k_{3},\ldots, k_{d})$ 
represent the momentum components perpendicular and parallel to the Fermi surface at $\pm K^*$, respectively. 
The momentum is rescaled such 
that the absolute value of the Fermi velocity,
and the quadratic curvature of the Fermi surface at $\pm K^*$, 
are equal to one.
Because the Fermi surface is locally parabolic, it is natural
to set the scaling dimension of $k_1$ (${\vec L}_{(k)}$) to be $1$ ($1/2$).
$H( {\vec L}_{(k)}^2 ) = \sum_{n=3}^\infty \sum_{i_1,..,i_n=2}^d \frac{c_{i_1,..,i_n}}{k_F^{\frac{n-2}{2}}} k_{i_1}..k_{i_n}$ 
denotes cubic and higher order terms in ${\vec L}_{(k)}$,
where $k_F$ is a scale of dimension $1$. 
The range of ${\vec L}_{(k)}$ in $\int dk$
is set by the size of the Fermi surface,
which is of order $k_F^{1/2}$ in this coordinate system. 
The Yukawa coupling has been also expanded around $\pm K^*$,
and only the momentum independent part is kept.
For the Ising-nematic quantum critical point, which we consider in this paper,
we have $e_+=e_-$.

\begin{figure}[h!]
\begin{center}
\subfigure[]{\includegraphics[scale=0.33]{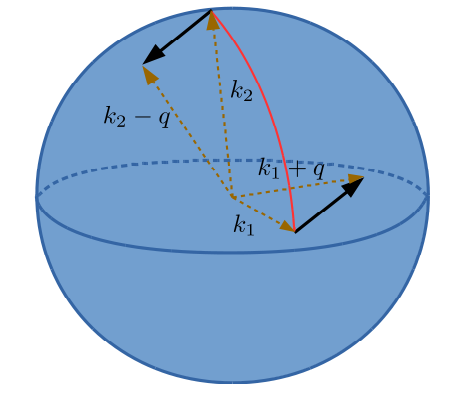} 
\label{fig:patch1}}
\subfigure[]{\includegraphics[scale=0.33]{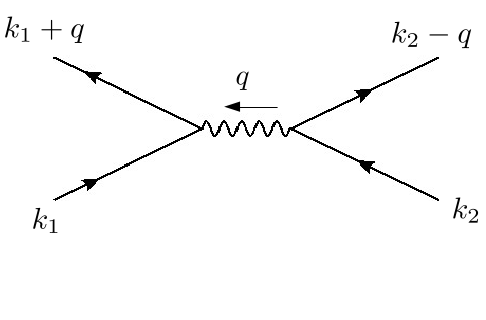} 
\label{fig:patch2}}
\end{center}
\caption{For $m>1$, any two points 
on the Fermi surface have at least $(m-1)$ common tangent vectors.
}
\label{fig:patch}
\end{figure}

The action is reminiscent of patch theories that have been used to describe non-Fermi liquids with one-dimensional Fermi surface. 
However, there is also an important difference between the theory 
with general $m>1$ and the one with $m=1$.
For $m>1$, any two points 
on the Fermi surface have at least $(m-1)$ common tangent vectors,
and the whole Fermi surface remains strongly coupled at low energies. 
Fig.~\ref{fig:patch} illustrates this point for a spherical Fermi surface embedded in a three-dimensional momentum space. 
Suppose that two fermions at general momenta $k_1, k_2$ are scattered to $k_1+q, k_2-q$
by exchanging a boson with small momentum $q$.
Because $q$ can be tangential to the Fermi surface  both at $k_1$ and $k_2$ for $m>1$,
the fermions can stay near the Fermi surface before and after scattering.
Therefore any two fermions on the Fermi surface remain strongly coupled in the low energy limit
even though the processes with large momentum exchanges are suppressed.
For $m=1$, such low-energy scattering is not present except for the two anti-podal points.
Because of the coupling that is global in the momentum space,
the theory with $m>1$ is ill-defined in the $k_F \rightarrow \infty$ limit
unlike the case with $m=1$.
In other words, low-energy (IR) observables, 
such as the fermion and boson Green's functions
near $k=0$, can not be defined until global properties, 
such as the size and shape, of the Fermi surface
are specified at large momenta (UV) for $m>1$.
In Fermi liquids, this UV/IR mixing is encoded
in the Landau parameters which are non-local in the momentum space.
It is our goal to understand 
consequences of the UV/IR mixing 
in non-Fermi liquid states with $m>1$.

The scale $k_F$ in $H( {\vec L}_{(k)}^2 )$ provides
a large momentum cut-off along the directions parallel to the Fermi surface.
Although irrelevant by power counting,
it is crucial to include the higher order terms
to keep the information that the Fermi surface is compact.
In principle, one has to keep an infinite set of independent 
parameters $c_{i_1,..,i_n}$ 
that encodes the precise shape of the Fermi surface away from $K^*$.
Here we consider a simplified `UV regularization'
which keeps the essential physics of the higher-order terms, 
but is simple enough to be amenable to an analytic treatment.
Specifically, we consider a regularized kinetic term
\bqa
\sum_{s,j} \int dk \, 
\psi_{s,j}^\dagger (k)
\Bigl[ 
 i k_0   +  s  k_{d-m} +   {\vec L}_{(k)}^2  \Bigr] \psi_{s,j}(k) \, 
\exp \Big \lbrace \frac {{\vec{L}}_{(k)}^2}  { k_F } \Big \rbrace. \nn
\label{kf}
\eqa
Here we keep the dispersion parabolic,
but the exponential factor effectively makes
the size of the Fermi surface finite
by damping the propagation of fermions 
with $|{\vec{L}}_{(k)}| > k_F^{1/2}$,
as is illustrated in Fig.~\ref{fig:patches}.

In order to control the Yukawa coupling and the strength of UV/IR mixing independently,
we tune both the dimension \cite{Chakravarty,Fit,Torroba} and the co-dimension of the Fermi surface \cite{senshank,Lee-Dalid,shouvik2}.
To keep the analyticity of the theory in momentum space 
(locality in real space) with general co-dimensions, 
we introduce a spinor \cite{Lee-Dalid,shouvik2}
$ \Psi_j^T(k) = \left( 
\psi_{+,j}(k),
\psi_{-,j}^\dagger(-k)
\right)$ 
to write an action that describes an $m$-dimensional Fermi surface
embedded in the $d$-dimensional momentum space:
\begin{eqnarray}
\label{act5}
S  &=&   \sum_{j} \int dk \bar \Psi_j(k)
\Bigl[ 
i \vec \Gamma \cdot \vec K  
+ i \gamma_{d-m} \, \delta_k \Bigr] \Psi_{j}(k) \, \exp \Big \lbrace \frac {{\vec{L}}_{(k)}^2}  { \mu \, {\tilde{k}}_F } \Big \rbrace \nonumber\\
&+&
\frac{1}{2} \int  dk ~
  {\vec{L}}_{(k)}^2\,  \phi(-k) \, \phi(k) \nonumber \\
 &+&     \frac{i \, e \, \mu^{x/2} }{\sqrt{N}}  \sum_{j}  
\int dk dq  \,
\phi(q) \, \bar \Psi_{j}(k+q)\,  \gamma_{d-m} \Psi_{j}(k) \,.
\end{eqnarray}
Here, $\vec K ~\equiv ~(k_0, k_1,\ldots, k_{d-m-1})$ includes
the frequency and the first $(d-m-1)$ components 
of  the $d$-dimensional momentum vector, ${\vec L}_{(k)} ~\equiv~ (k_{d-m+1}, \ldots, k_{d})$ and $\delta_k =  k_{d-m}+ {\vec{L}}_{(k)}^2$.
In the $d$-dimensional momentum space,
$k_1,..,k_{d-m}$ (${\vec L}_{(k)}$) represent(s) the
$(d-m)$ ($m$) directions perpendicular (tangential) to the Fermi surface.
$\vec \Gamma \equiv (\gamma_0, \gamma_1,\ldots, \gamma_{d-m-1})$ represents the gamma matrices associated with $\vec K$.
Since we are interested in co-dimension $1 \leq d-m \leq 2$, 
we consider only $2 \times 2$ gamma matrices with
$\gamma_0= \sigma_y , \, \gamma_{d-m} = \sigma_x$
and $\bar \Psi \equiv \Psi^\dagger \gamma_0$. 

The leading terms in the quadratic action in Eq.~(\ref{act5})
are invariant under the scale transformations:
\begin{eqnarray}
\vec K & = & \frac{\vec K'}{b} \,, \quad k_{d-m} =\frac{k_{d-m}'}{b} \,,\quad {\vec{L}}_{(k)} = \frac{{\vec{L}}_{(k)}'}{\sqrt{b}} \,,\label{s1} \nn
\Psi_j (k) & = & b^{ \frac{2 d + 4-m}{4}} \Psi'_j (k') \,, \quad \phi (k) = b^{\frac{2 d + 4-m}{4}} \phi'(k')\,. 
\label{s4} 
 \end{eqnarray}
In the quadratic action of the boson, only ${\vec{L}}_{(k)}^2 \, \phi^* (k)\phi(k)$ 
is kept, because $|\vec K|^2 + k_{d-m}^2$ 
is irrelevant under the scaling 
where $k_0,k_1,..,k_{d-m}$ have dimension $1$
and $k_{d-m+1},..,k_d$ have dimension $1/2$.
In the presence of the $(m+1)$-dimensional rotational symmetry,
all components of $k_{d-m}, ..., k_d$ should be equivalent.
The reason why $k_{d-m}$ is treated differently from ${\vec{L}}_{(k)} = (k_{d-m+1},..,k_d)$
is because bosons that are strongly coupled to the fermions around $\pm K^*$ have momentum 
$ |{\vec{L}}_{(k)}| >> k_{d-m}$.
Therefore we ignore the dependence on $k_{d-m}$ in the boson kinetic term 
in the effective theory that describes the regions around $\pm K^*$.

The scaling dimension of the Yukawa coupling is $x/2\,,$ where
\beq
\label{e-dim}
x = \frac{4+m-2d} {2} \,.
\eeq
Here, $e$ is dimensionless and $\mu$ is a mass scale.
We also define a dimensionless parameter for the Fermi momentum, 
$ {\tilde {k}}_F = k_F/\mu$.
The spinor has the energy dispersion with two bands,
$E_k = \pm \sqrt{ \sum_{i=1}^{(d-m-1)} k_i^2 + \delta_k^2  }$,
which gives an $m$-dimensionsal Fermi surface 
embedded in the $d$-dimensional momentum space,
defined by the $d-m$ equations:
$k_i = 0$ for $i=\lbrace 1,\ldots,d-m-1 \rbrace$
and ${ k}_{d-m}  = - {\vec{ L}}_{(k)}^2$.

\begin{figure}
\includegraphics[scale=0.42]{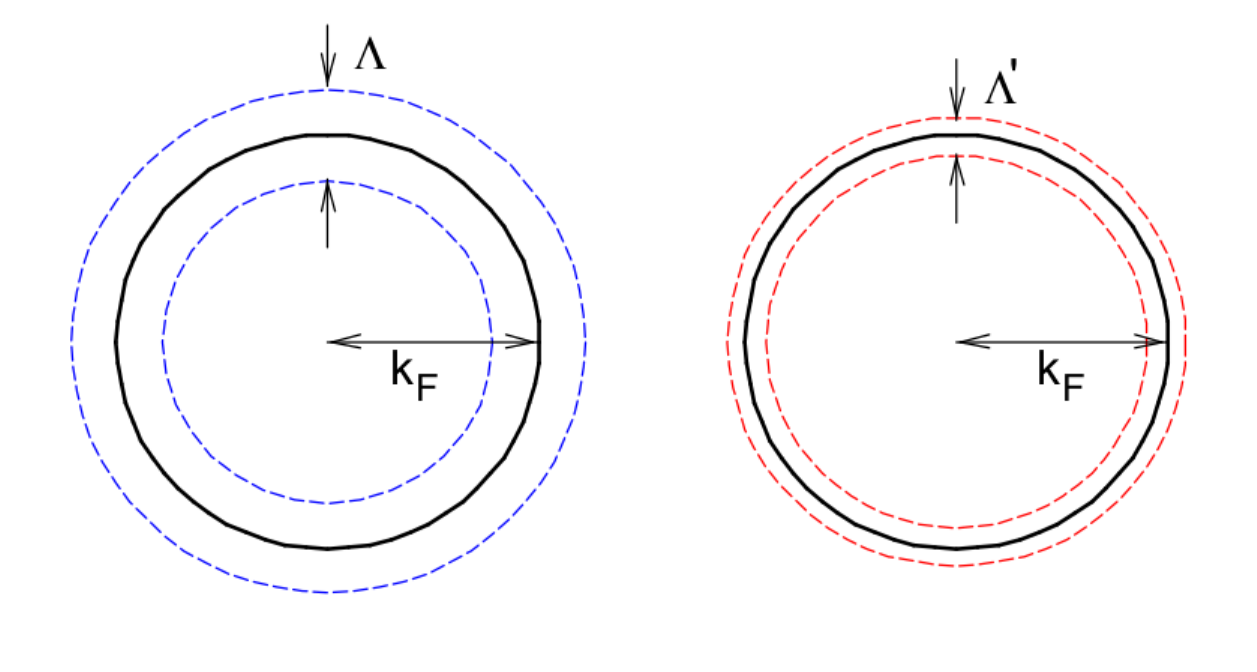} 
\caption{
As the high-energy modes away from the Fermi surface are integrated out,
the ratio $k_F/\Lambda$ grows. 
$k_F$ is the size of the Fermi surface 
and $\Lambda$ is the energy cut-off perpendicular to the Fermi surface.
For $m>1$, the Green's functions are singular in the $k_F/\Lambda \rightarrow \infty$ limit,
which results in the UV/IR mixing.
}
\label{fig:FS2}
\end{figure}

Besides $k_F$ and $e$, the theory implicitly 
has a UV cut-off for $\vec K$ and $k_{d-m}$,
which we denote as $\Lambda$.
It is natural to choose $\Lambda = \mu$,
and the theory has two important dimensionless parameters : $e$, $\tilde k_F = k_F / \Lambda$.
If $k$ is the typical energy 
at which we probe the system, 
the limit of interest is $k << \Lambda << k_F$. 
This is because  $\Lambda$ sets the largest energy (equivalently, momentum perpendicular to the Fermi surface) 
fermions can have, whereas $k_F$ sets the size of the Fermi surface.
We will consider the renormalization group flow generated 
by changing $\Lambda$ and requiring that low-energy observables are independent of it.
This is equivalent to a coarse-graining procedure of integrating out
high-energy modes away from Fermi surface.
Because the zero energy modes are not integrated out,
$k_F/\Lambda$ keeps increasing  in the coarse graining procedure.
We treat $k_F$ as a dimensionful coupling constant 
that flows to infinity in the low-energy limit.
Physically, this describes the fact that 
the size of the Fermi surface measured in the unit of
the thickness of the thin shell around the Fermi surface
diverges in the low-energy limit.
This is illustrated in Fig. \ref{fig:FS2}.

\section{Dimensional Regularization}
\label{dr}

To access perturbative non-Fermi liquids, 
we fix $m$ and tune $d$ towards a critical dimension, 
at which quantum corrections depend logarithmically on $\Lambda$ 
within the range $\Lambda << k_F$.
The Yukawa coupling is dimensionless
at  
\beq
d_c'(m) =\frac{4 + m}{2} \,.
\eeq
However, it turns out that this is not the actual upper critical dimension 
at which the quantum corrections diverge logarithmically in $\Lambda$.
The shift of the upper critical dimension is a 
sign that $k_F$ enters the low-energy physics
in a way that is singular in the large $k_F$ limit,  
resulting in UV/IR mixing.

\begin{figure}[h!]
\begin{center}
\subfigure[]{\includegraphics[scale=0.3]{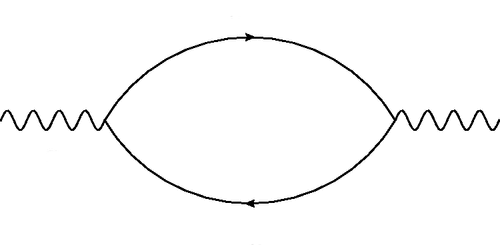} 
\label{fig:bos}} \\
\subfigure[]{\includegraphics[scale=0.3]{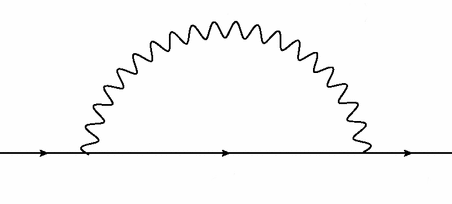} 
\label{fig:ferm}} \\
\subfigure[]{\includegraphics[scale=0.3]{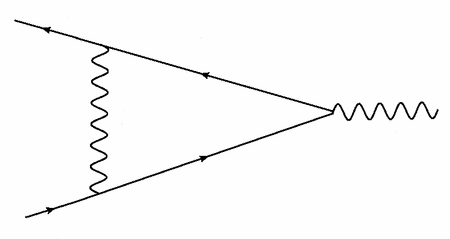} 
\label{fig:vert}} \\
\end{center}
\caption{The one-loop diagrams
for the boson self-energy (a),
the fermion self-energy (b),
and the vertex correction (c).
Solid lines represent the bare fermion propagator,
whereas wiggly lines in (b) and (c) represent
the dressed boson propagator 
which includes the one-loop self-energy in (a).
}
\label{fig:1loop}
\end{figure}

In order to identify the actual upper critical dimension,
we consider the one-loop quantum corrections.
Since the bare boson propagator is independent of $k_{0},..,k_{d-m}$,
the loop integrations involving it are ill-defined,
unless one resums a series of diagrams
that provides a non-trivial dispersion along those directions.
This amounts to rearranging the perturbative expansion 
such that the one-loop boson self-energy is included 
at the `zero'-th order.
The dressed boson propagator, which includes the one-loop self-energy
(Fig. \ref{fig:bos}), is given by
\beq
\label{babos}
D_1(k) = \frac{1}{ {\vec{L}}_{(k)}^2 +\beta_{d} \, e^2 \, \mu^{x}
\displaystyle\frac{ ( \mu \, {\tilde{k}}_F )^{ \frac{m-1}{2}}  \, |\vec K|^{d-m}}{ |\vec{L}_{(k)}| } } \,,
\eeq
to the leading order in $k/k_F$,
for $|\vec K|^2/|\vec L_{(k)}|^2, ~\delta_k^2/|\vec L_{(k)}|^2 << k_F$. Here
$\beta_d = \frac{  \Gamma^2 (\frac{d-m+1} {2})}
{2^{\frac{2d+m-1}{2}}  \pi^{\frac{d-1}{2}}  | \cos \lbrace  \frac{\pi (d-m+1)} {2} \rbrace |   \Gamma(\frac{d-m}{2}) \Gamma (d-m+1)}$
is a parameter of the theory that 
depends on the shape of the Fermi surface.
See Appendix~\ref{app:oneloop} for the derivation of the one-loop self-energies.
Since the boson propagator depends on $e$, 
the higher-loop diagrams are no longer suppressed by $e^2$,
but by a fractional power of $e$ \cite{Lee-Dalid}.
Moreover, the boson self-energy diverges
in the $k_F \rightarrow \infty$ limit for $m>1$.
This is due to the fact that the Landau damping gets stronger 
for a system with a larger Fermi surface,
as the boson can decay into particle-hole excitations
that encompass the entire Fermi surface for $m>1$.
This is in contrast with the case for $m=1$, 
where a low-energy boson with a given momentum can decay into particle-hole excitations only near the 
isolated patches whose tangent vectors are parallel
to that momentum.
For $m=1$,  $k_F$ drops out in Eq. (\ref{babos}), 
which indicates the absence of UV/IR mixing.
Eq. (\ref{babos}) is valid when there exists at least one direction that is tangential to the Fermi surface,
and it should not be extended to the cases with $m<1$ for which the conventional quantum field theories work well.
From now on, we will focus on the case with $m>1$.


The apparent lack of rotational symmetry in the space of $k_{d-m},..,k_d$ in Eq. (\ref{babos}) is 
because the expression is valid only for the boson whose momentum is almost tangential to the Fermi surface at points $\pm K^*$
as is discussed below Eq. (\ref{s4}).
For the boson propagator with general momentum,
$ |{\vec L}_{(k)} | $ in Eq.~(\ref{babos}) should be replaced by $\sqrt{ k_{d-m}^2 + ... + k_d^2 } $
in the presence of $(m+1)$-dimensional rotational symmetry.
This is because for any given boson momentum $k$, one can always find a point on the Fermi surface where $ k $ is tangential to the Fermi surface. 
If one chooses a coordinate system where $ k_{d-m}=0 $,  the boson self-energy takes the exactly same form as in Eq.~(\ref{babos}). 
Since we can do this for any $k$, the boson propagator with general momentum should be independent of the direction in the space of $ (k_{d-m}, \cdots ,k_d) $. 
In the following,  we will use the expression in  Eq. (\ref{babos}) because 
we are only interested in describing the Fermi surface near $\pm K^*$.
The bosons that are strongly coupled to that region have momentum $k_{d-m} << |{\vec L}_{(k)} | $,
and  $\sqrt{ k_{d-m}^2 + ... + k_d^2 } $ is reduced to $ |{\vec L}_{(k)} | $. 

\begin{figure}
\includegraphics[width= 0.48 \textwidth]{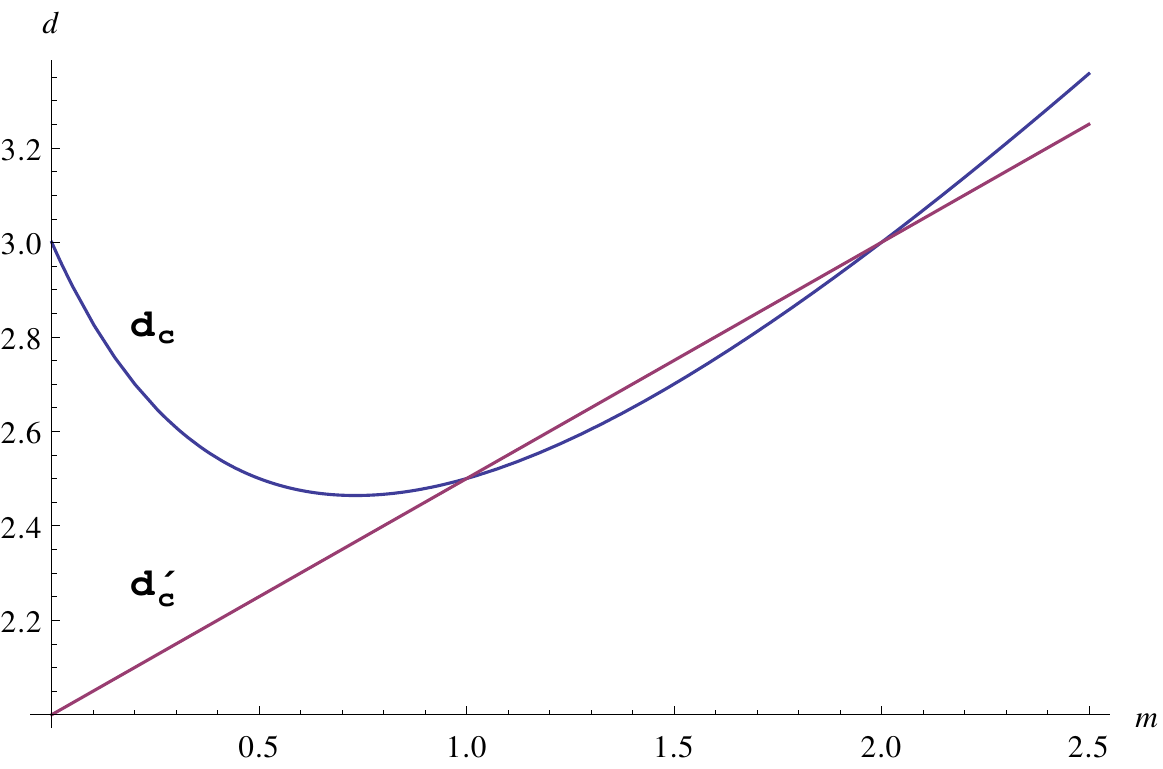}
\caption{The plots for $d_c$ and $d_c'$ as functions of $m$.}
\label{fig:d-dc}
\end{figure}


By using the dressed boson propagator,
we have computed the one-loop fermion self-energy $\Sigma_1 (q)$ (Fig.~\ref{fig:ferm})  in Appendix~\ref{app:oneloopferm}.
This blows up logarithmically 
in $\Lambda$ at a new critical dimension
\beq
d_c(m) = m + \frac{3}{m+1} \,,
\eeq
which is smaller than $d_c^{'}$ for general $1 <m<2$.
Fig.~\ref{fig:d-dc} shows the plots of $ d_c $ and $d_c'$ as functions of $m$.
Now we consider the space dimension $d=d_c(m) - \epsilon$.
In the dimensional regularization scheme, 
the logarithmic divergence in $\Lambda$
turns into a pole in $1/\epsilon$:
\beq
\label{sigma4}
\Sigma_1 (q) = 
\left( - \frac{e^{ 2 (m+1 ) /3}  } {{\tilde{k}}_F ^{ \frac{(m-1) (2-m)} {6}}N} 
\frac{u_1}{\epsilon} 
+ \mbox{ finite terms}
\right)
(i \vec \Gamma \cdot \vec Q) \,,
\eeq
to the leading order in $q/k_F$, where
$u_1 = \frac{ 1   }
{ \pi^{\frac{m-2}{2}}
(4 \pi)^{\frac{3}{2(m+1)}} 2^{m-1}   | \sin \lbrace (m+1) \pi/3 \rbrace |  \beta_{d}^{\frac{2-m}{3}}  (m+1) }$
$\times \frac{  \Gamma (  \frac{m+4}{2(m+1)})  } 
{  \Gamma{(m/2)}  \Gamma(\frac{2-m}{2(m+1)})  \Gamma (  \frac{2m+5}{2(m+1)}) }$.
The one-loop vertex correction in Fig. \ref{fig:vert}
vanishes due to a Ward identity \cite{Lee-Dalid}.

It is noted that one can tune the dimension of Fermi surface from $ m = 1$ to $ m=2 $ while keeping $\epsilon$ small, 
thus providing a controlled description for any $m$ between $1$ and $2$.
This is possible because we are tuning $m$ and $d$ independently.
For a given $m$, we tune $d$ such that $\epsilon =  d_c(m) - d $ is small.
To remove the UV divergences in the $\epsilon \rightarrow 0$ limit,
we add counterterms using the minimal subtraction scheme.
Adding the counterterms to the original action,
we obtain the renormalized action
which gives the finite quantum effective action:
\begin{widetext}
\begin{eqnarray}\label{act7}
S_{ren} & = &  \sum_{j} \int d k_B
\, \bar \Psi_{Bj}(k_B)
\Bigl[ 
i  \vec \Gamma \cdot \vec K_B + 
i \gamma_{d-m} \delta_{k_B}  \Bigr] \Psi_{Bj}(k_B) \, \exp \Big \lbrace \frac {{\vec{L}}_{(k),B}^2}  {  k_{F,B} } \Big \rbrace
\nonumber\\
&+&
\frac{1}{2} \int  d k_B \,
 {\vec{L}}_{(k)}^2\,  \phi_B(-k_B) \, \phi_B(k_B) \nonumber \\
 &+&     \frac{i \, e_B }{\sqrt{N}}  \sum_{j}  
\int d k_B \, d q_B \,
\phi_B(q_B) \, \bar \Psi_{Bj}(k_B+q_B) 
\, \gamma_{d-m} \Psi_{Bj}(k_B) \, ,
\end{eqnarray}
\end{widetext}
where
\bqa
\vec K & = & \frac{Z_2}{Z_1} \vec K_B \, , \quad k_{d-m} = k_{B, d-m} \, , \quad {\vec{L}}_{(k)} = {\vec{L}}_{(k), B} \,, \nn
\Psi(k) & = &  Z_\Psi^{-1/2} \,\Psi_B(k_B)\,, \quad \phi(k) = Z_\phi^{-1/2} \phi_B(k_B)\,, \nn
e_B & = & Z_3^{-1/2} \left( \frac{Z_2}{Z_1} \right)^{(d-m)/2} \mu^{ x/2} \, e \,,\quad   k_F =\mu \, {\tilde{k}}_F \,,
\eqa
with 
\beq
Z_\Psi =  Z_2 \left( \frac{Z_2}{Z_1} \right)^{(d-m)}\,, \quad
Z_\phi = Z_3 \left( \frac{Z_2}{Z_1} \right)^{(d-m)}\,.
\eeq
The subscript ``B'' denotes the bare quantitites.
To the one-loop order, we have $Z_n = 1 + \frac{Z_{n,1}}{\epsilon}$ with
\bqa
Z_{1,1} & = &  - \frac{e^{ 2 (m+1 ) /3}  \, u_1 } {{\tilde{k}}_F ^{ \frac{(m-1) (2-m)} {6}}N }  \,, \nn
Z_{2,1} & = & 0 \,, \nn
Z_{3,1} & = & 0 \,. \nn
\eqa

The one-loop beta functions, that dictate the flow of $\tilde k_F$ and $e$
with the increasing logarithmic length scale $l$, 
are given by
\bqa
 \frac{d \tilde k_F }{dl}  &=&  \tilde k_F  , \label{eq:bkf} \\
 \frac{d e}{d l}  &=&  
\left[ \frac{\epsilon}{2} + \frac{(m-1) \, (2-m)}{4 \, (m+1)} \right]  e 
- \frac{ u_1 \, {\tilde e} }{2N} \, e \,,
\label{eq:be}
\eqa
with
\beq
{\tilde e} \equiv 
\frac{e^{ 2 (m+1 ) /3}  } {{\tilde{k}}_F ^{ \frac{(m-1) (2-m)} {6}}} \,.
\label{eq:eeff}
\eeq 
$\tilde k_F$ increases under the RG flow
because the size of the Fermi surface, measured
in the unit of the floating energy scale $\mu  \exp({-l})$,
increases at low energies.
The first term in the beta function of $e$ indicates
that $e$ remains strictly relevant at $d=d_c(m)$ for $1 < m < 2$.
However, the form of the loop correction (the second term) implies that
the higher order corrections are controlled not by $e$, 
but by an effective coupling ${\tilde e}$.
This can be also checked for higher-order diagrams.
The beta function for ${\tilde e}$, which no longer contains $\tilde k_F$, is given by
\beq
\label{bstable}
\frac{d {\tilde e} }{dl}  = \frac{(m+1)\, \epsilon}{3} \, {\tilde e} \,
- \frac{ (m+1)\,  u_1} {3N} \, {\tilde e}^2 \,,
\eeq
to order ${\tilde e}^2$.


Eq.~(\ref{bstable}) shows that the effective coupling flows to an IR stable fixed point at
\beq
\label{efp}
{\tilde e}^*= \frac{N \epsilon} {u_1}  + \mathcal{O}(\epsilon^2)\,.
\eeq
For small $\epsilon$, the interacting fixed point is perturbatively accessible
despite the fact that the scaling dimension of the bare coupling $e$ stays positive 
in the $\epsilon \rightarrow 0$ limit for $1 < m < 2$.
Although $e$ grows at low energies, 
the increase of the bare coupling is compensated by 
the Landau damping, which also increases
with the effective size of the Fermi surface.
The competition between 
the interaction and the Landau damping
makes the effective coupling marginal at the new critical dimension $d_c$. 
It is interesting that $k_F$ drops from the effective coupling 
not only for $m=1$ but also for $m=2$. For the latter case, the $k_F$ dependence in the Landau damping
cancels out the $k_F$ dependence from the phase space of intermediate states in Fig.~\ref{fig:ferm}. 
However, the UV/IR mixing is present for all $m>1$ 
because the Landau damping diverges in the large $k_F$ limit.

\section{Renormalization Group Equations}
\label{rg}
The renormalized Green's functions, defined by
$\Bigl< \phi( k_{1} ) .. \phi(k_{{n_\phi} })$
$\Psi( k_{{n_\phi} +1} ).. \Psi(k_{{n_\phi}+n_\psi })$
$ \bar \Psi( k_{n_{\phi}+n_\psi+1} )  .. \bar \Psi(k_{n_\phi+2 n_\psi} ) \Bigr> $
$ = G^{({n_\psi} ,{n_\psi} , {n_\phi} )}( \{ k_i \}; \tilde e, \tilde k_F , \mu )$
$\delta^{d+1} \left( \sum_{i=1}^{{n_\phi+n_\psi} } k_i  - \sum_{j={n_\phi} + n_\psi +1}^{2 {n_\psi} + {n_\phi}} k_j \right)$,
satisfy the RG equations
\begin{widetext}
\bqa
\Bigg\{
- \sum_{i=1}^{2 {n_\psi}  + {n_\phi} } \left(
z \, \vec K_i \cdot \nabla_{K_i}
+ k_{i, d-m}  \frac{\partial}{\partial k_{i,d-m}}
+ \frac{ \vec{L}_{(k_i)} } {2}   \cdot  \nabla_{{L}_{(k_i)}}
\right)  
- \frac{d \tilde k_F}{dl}  \frac{\partial} {\partial \tilde k_F}  
-  \frac{d {\tilde e}}{dl}   \frac{\partial}{\partial {\tilde e}} 
 + 2  \, {n_\psi}  \left ( -  \frac{2 \, d_c - 2 \, \epsilon + 4 -m }{4} +  \eta_\psi \right)  && \nn
+  {n_\phi} \left ( -  \frac{2 \, d_c - 2 \, \epsilon +4 -m }{4} +  \eta_\phi \right) 
+  d_c - \epsilon+1 - \frac{m}{2}
+ (d_c - \epsilon - m) (z-1)
\Bigg\}
\, G^{({n_\psi} , {n_\psi}  ,{n_\phi})}(\{ k_i \}; {\tilde e}, \tilde k_F , \mu ) 
=0 \,. ~~~~~~~~~ &&
\label{RGeq}
\eqa
\end{widetext}
Here the dynamical critical exponent $z$ and the anomalous dimensions are given by
\bqa
 z^* =  \frac{3}{3 - (m+1) \epsilon } \,, ~~~
\eta_\psi^*=\eta_\phi^*  = -\frac{ \epsilon}{2} \,,
\eqa
to the one-loop order at the fixed point.
It is remarkable that the exponents are insensitive to
the details of the Fermi surface (such as $\beta_d$)
despite the fact
that patch scaling is violated by $k_F$. 
This vindicates our use of the exponential cut-off scheme
in Eq. (\ref{kf}) which captures the compactness of the Fermi surface
in a minimal way without including the details of the shape.
The finite anomalous dimensions are the result of the dynamical balance
between the two strongly relevant couplings, $e$ and $k_F$.
This is opposite to the case where finite anomalous dimensions result from
a balance between two irrelevant `couplings' \cite{shouvik2}.

From this, one can write down the general scaling form of the two-point functions 
at the IR fixed point as:
\bqa
G^{(0,0,2)} &= & \frac{1}{ \left (  \vec{L}_{(k)}^2 \right )^{ 2 \Delta_\phi } } \,
f_D \left( 
\frac{| \vec K |^{1/z^*} }{  \vec{L}_{(k)}^2 } ,  \frac{k_{d-m} }{k_F} ,  \frac{  \vec{L}_{(k)}^2 } {k_F}
\right),  \label{Dk0} \nn
G^{(1,1,0)} & = &  \frac{1}{ | \delta_k |^{ 2 \Delta_\psi} } \,
f_G 
\left( \frac{ | \vec K |^{1/z^*} }{\delta_k} , \frac{\delta_k }{k_F} ,  \frac{  \vec{L}_{(k)}^2 } {k_F}
 \right) , \label{Gk0}
\eqa
where 
$2 \Delta_\phi = 1 - (z^*-1) \left (\frac{3}{m+1}-\epsilon \right) - 2 \eta^*_\phi = 1+O(\epsilon^2)$,
$2 \Delta_\psi = 1 - (z^*-1) \left (\frac{3}{m+1}-\epsilon \right) - 2 \eta^*_\psi = 1+O(\epsilon^2)$.
To the one-loop order, the universal scaling functions are given by
\bqa
f_D( X,Y,Z) & = & \left[
 1 +
\beta_d \, {\tilde e}^{\frac{3}{m+1}}
X^{\frac{3}{m+1} }
Z^{-\frac{3(m-1)}{2(m+1)}}
\right]^{-1}, \\
f_G( X,Y,Z) & = & -i \left[ C \, ( {\vec  \Gamma} \cdot \hat {\vec K} ) \, X   + \gamma_{d-m} \right]^{-1} ,
\eqa
in the $ Y, Z \rightarrow 0$ limit with fixed $X$. 
Here 
$C = 
\mu^{ \frac {m+1}{3} \epsilon  }
\Big \lbrace
1 - \frac{(m+1) \, \gamma \, \epsilon } {6}
\Big \rbrace,
$ and $\gamma$ is the Euler-Mascheroni constant. 
It is noted that $f_D$ has a singular dependence on $ Z $ 
in the small $  Z$ limit.
%
%
The absence of the sliding symmetry 
makes the fermion Green\rq{}s function depend on $\delta_k$ and $\vec L_{(k)}$ separately in general.

\section{Physical relevance of the dimensional expansion}
\label{dimexp}

The motivation for the dimensional expansion is to understand the stark difference in the behaviors of non-Fermi liquids in two and three dimensions. Non-Fermi liquids in two dimensions can at most have one-dimensional Fermi surface,
for which $k_F$ does not play an important role in the low-energy scaling.
On the other hands, $k_F$ enters as an important scale in three dimensions due to UV/IR mixing.
Our goal is to understand how this transition occurs in a systematic way
by tuning the dimension and co-dimension continuously.
Although systems with non-integer dimensions are unphysical by themselves,
they provide an insight on how dimension and co-dimension contribute to different behaviors
of metals in the physical dimensions.

In physical dimension with $d=3,m=2$, $k_F$ drops out in the effective coupling.
However, the non-trivial UV/IR mixing still manifests itself in the dispersions of fermion and boson. 
Near $d=3$, fermion has the dispersion $k_0 \sim k_x + L_k^2$, whereas the boson has the scaling $k_0 \sim L_k^3$ upto small corrections.
Our theory provides a scaling consistent both for boson and fermion by including $k_F$ as a dimensionful parameter of the theory.
Boson and fermion can have  different effective dynamical critical exponents at the scale-invariant fixed point, 
because the difference of the dynamical critical exponents is compensated by $k_F$.
This is in contrast to the case with $m=1$, where the dispersions of the boson and fermion obey the same scaling behavior.
UV/IR mixing also plays an important role in suppressing higher-loop quantum corrections for $m>1$.
In the following section, we will examine the effect of $k_F$ in higher-loop diagrams.

\section{Expansion Parameter}
\label{ep}

The present work is an extension of the early work \cite{Lee-Dalid} which provides a controlled expansion for non-Fermi liquids supporting one-dimensional Fermi surface ($m=1$).
For $m=1$, it was explicitly shown that two and three-loop diagrams are suppressed by positive powers of $\tilde e$, which is of the order $\epsilon$. 
There also exists a general argument outlining why higher-loop diagrams are systematically suppressed by higher powers of $\tilde e$. 
The expansion here is different from an expansion in powers of $1/N$, 
and it does not suffer from the proliferation of planar diagrams in the $1/N$ expansion \cite{SSLee,metlsach}. 
Due to the addition of the extra co-dimensions, the density of states is suppressed at low energies. 
The weaker IR singularity allows one to have a controlled expansion for a sufficiently small $\epsilon$. 
In the present paper, we have generalized this to the cases where the dimension of Fermi surface is greater than one ($m>1$).
For $m>1$, the suppression of higher-loop diagrams by positive powers of $\tilde e$ is unchanged. 
The difference for $m>1$ is the presence of an additional scale $k_F$.

\begin{figure}
\includegraphics[scale=0.35]{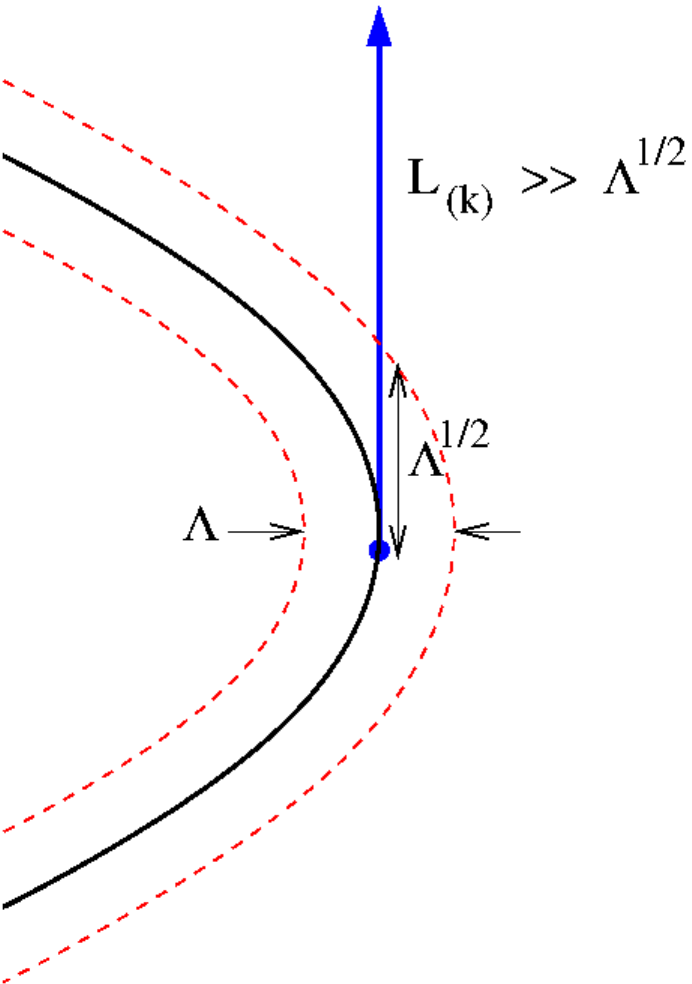} 
\caption{
A two-dimensional slice of an $m$-dimensional Fermi surface.
The typical momentum carried by a boson
is proportional to 
${\tilde \alpha}^{1/3} \Lambda^{\frac{d-m}{3}} \sim {\tilde e}^{\frac{1}{m+1}} \left( \frac{k_F}{\Lambda} \right)^{\frac{m-1}{2(m+1)}} \Lambda^{1/2}$.
For $m>1$, this momentum is much larger than $\Lambda^{1/2}$
in the low-energy limit.
As a result, the momentum transferred from a boson
takes a fermion near the Fermi surface
outside the thin sell of the UV cut-off. 
This leads to a suppression of the virtual particle-hole excitations
by powers of $\Lambda/k_F$ for $m>1$.
}
\label{fig:crossover}
\end{figure}

In order to estimate the magnitudes of higher-loop corrections, 
we first discuss an interplay between $k_F$ and $\Lambda$
that plays an important role for $m>1$.
Suppose $k = ( {\bf K}, k_{d-m}, \vec L_{(k)} )$ denotes the momentum that flows through a boson propagator 
within a two-loop or higher-loop diagram.
When $|{\bf K}|$ is order of $\Lambda$,
the typical momentum carried by a boson along the tangential direction of the Fermi surface is given by
\beq
\label{cross}
 |{\vec{L}}_{(k)} |^3 \sim 
\tilde \alpha  \, \Lambda^{d-m} 
\,,
\eeq
where
\beq 
 \tilde{\alpha} =   \beta_{d} \, e^2 \, \mu^{x} \,( \mu \, {\tilde{k}}_F )^{ \frac{m-1}{2}}.
 \label{alpha}
\eeq
This can be seen from the form of the boson propagator, which is given in Eq.~(\ref{babos}).
If $ \left( \tilde \alpha  \, \Lambda^{d-m}  \right)^{1/3} >> \Lambda^{1/2} $,
the momentum imparted from the boson to fermion is much larger than $\Lambda^{1/2}$
as is illustrated in Fig. \ref{fig:crossover}.
In this case, the typical energy of virtual particle-hole excitations within the loop is much larger than $\Lambda$.
As a result, the loop contributions are suppressed by a power of $\Lambda/k_F$ at low energies.
On the other hand, there is no such suppression if $ \left( \tilde \alpha  \, \Lambda^{d-m}  \right)^{1/3} << \Lambda^{1/2} $.
The crossover is controlled by the dimensionless quantity,
\beq
\lambda_{\text{cross}} \equiv
 \tilde{e}^{2} \, 
\left ( \frac{  k_F }  {\Lambda  } \right )^{ m-1 } \,,
\label{crossover}
\eeq
which determines whether $ \left( \tilde \alpha  \, \Lambda^{d-m}  \right)^{1/3} >> \Lambda^{1/2} $ or 
$ \left( \tilde \alpha  \, \Lambda^{d-m}  \right)^{1/3} << \Lambda^{1/2} $.

For $m=1$, $k_F$-dependence drops out from everywhere. 
Since $\tilde{e} \sim \mathcal{O} (\epsilon)$ within the perturbative window, 
one always deals with the limit, 
\beq
\label{lim3}
\lambda_{\text{cross}} << 1
\,, \quad \mbox{for} \,\, m=1\,.
\eeq
The situation is different for $m>1$.
Unlike the case with $m=1$, 
the tangential momentum carried by the boson 
depends on both $\Lambda$ and $k_F$.
For a fixed value of $\tilde{e} \sim \mathcal{O} (\epsilon)$,
one is always in the limit of 
\beq
\label{lim1}
\lambda_{\text{cross}}
>> 1
 \,, \quad \mbox{for} \,\, m>1 \,,
\eeq
at sufficiently low energies.
This is because $k_F$ has a positive scaling dimension,
and $k_F / \Lambda$ flows to $\infty$ in the low-energy limit.
The crossover occurs at the energy scale $\Lambda \sim \tilde e^{\frac{2}{m-1}} k_F$. 
It is noted that there exists a large energy window for small $\epsilon$ and $(m-1)$, before 
the theory enters into the low-energy limit controlled by $\lambda_{cross} >> 1$.
There can be non-trivial quantum corrections from higher-loop diagrams in this intermediate energy scale. We postpone the detailed study of the intermediate scale to a future work.
Here we focus on the low-energy limit with $ \lambda_{\text{cross}} >> 1 $.
In this limit, higher-loop diagrams are suppressed by $k_F$ as was shown in an earlier work \cite{schafer}, for the special case of $m=2$ in three dimensions.

For general $m >1$ with $\lambda_{cross}>>1$, 
the two-loop self-energies for boson and fermion are given by
\beq
\Pi_2 (q)  
 \sim 
\frac { \tilde{e}^{\frac{m} {m+1}} } {  k_F^{\frac{m-1}{2\,(m+1)} } }
\frac{ | \vec Q |^{\frac{  m } {m+1} } 
}
{ N \, | \vec L_{(q)} |  }   \, \Pi_1 (q)  \,,
\eeq

\bqa
&& \Sigma_{2a} (q)  
\sim  
{\tilde e}^{\frac{ 2\,(m-1) } {m+1}}
\left( \frac {\Lambda }{ k_F} \right)^{\frac{ 2\,(m-1)}{ m+1 } } 
i \, \gamma_{d-m} \, \delta_q \,,
\eqa
to the leading order in $\Lambda/k_F$. 
See Appendix~\ref{app:twoloop} for the derivation of the results. 
Here $\Pi_2 (q) $ is the two-loop boson self-energy contributed from Fig.~\ref{fig:bos2}(a).
$\Sigma_{2a} (q)$ and $\Sigma_{2b} (q) $ are the fermion self-energy from Fig.~\ref{fig:ferm2}(a), 
which are proportional to $\gamma_{d-m} \, \delta_q$ and $( \vec \Gamma \cdot \vec Q ) $ respectively.
Other diagrams in Figs.~\ref{fig:bos2}(b)-(e) and \ref{fig:ferm2}(b)-(c)  do not contribute \cite{Lee-Dalid}.
The coefficients in the expressions for $\Sigma_{2 a, 2b} (q)$ vanish at $d-m=1$.
The vertex correction is related to the fermion self-energy through the Ward identity.

Compared to the one-loop self-energies, the two-loop corrections are suppressed 
not only by $\tilde e$ but also by powers of $\Lambda / k_F $. 
Because of the suppression by $1/k_F$, there is no logarithmic or higher-order divergence at the critical dimension.
As a result, the critical exponents are not modified by the two-loop diagrams in the $k_F \rightarrow \infty$ limit.
It is noted that the suppression by $\Lambda/k_F$ originates from 
the large Landau damping which suppresses quantum fluctuations at low energies.
Since the suppression is not specific to the two-loop diagrams,
we expect that all higher-loop diagrams are also suppressed by $\tilde e$ and $1/k_F$ in the low-energy limit.
We have checked this explicitly for some three-loop diagrams.

\section{\label{conclusion}Conclusion}

To summarize, we have extracted the scaling behaviour of non-Fermi liquids 
with a Fermi surface of general dimensions and co-dimensions 
based on a dimensional regularization scheme. 
For $m>1$, the low-energy physics becomes sensitive to the size of Fermi surface $k_F$,
which results in UV/IR mixing.
As a result, the upper critical dimension is shifted from the one predicted by the power-counting, and
the perturbative expansion is controlled by a combination of the Yukawa coupling and the Fermi momentum $k_F$.
By tuning the dimension below the upper critical dimension, 
we have shown that there exists a stable non-Fermi liquid fixed point
where both interaction and UV/IR mixing play crucial roles.
We have also shown that the critical exponents at the low-energy fixed point 
are not modified by the two-loop diagrams, due to the UV/IR mixing for $m>1$.
This is likely to be the case for all higher-loop diagrams as well.

So far we have not considered the four-fermion interaction $V$,
which has the tree-level scaling dimension $-d+1+m/2$.
However, scatterings in the pairing channel are enhanced
by the volume of the Fermi surface $\sim k_F^{m/2}$.
As a result, the effective coupling that dictates the potential instability, driven by the four-fermion interactions, is given by $\tilde V = V k_F^{m/2}$, 
which has an enhanced scaling dimension $-d+1+m$.
$\tilde V$ is marginal at the tree-level for co-dimension $d-m=1$.
For a co-dimension $d-m>1$, there is no perturbative pairing instability for a sufficiently small $\epsilon = d_c-d$.
When $d-m -1 \lesssim \epsilon$ with $ d_c - d \sim \epsilon$, 
the interaction plays an important role for the pairing instability~\cite{ips-sudip1,ips-sudip2,Max,ips-sc}.


\begin{acknowledgments}
We thank Ganapathy Baskaran, Liam Fitzpatrick, Shamit Kachru, Max Metlitski, Sri Raghu,   
and Subir Sachdev for stimulating discussions.
The research was supported by NSERC, ERA and the Templeton Foundation.
Research at the Perimeter Institute is supported in part by the Government of Canada 
through Industry Canada, and by the Province of Ontario through the Ministry of Research and Information.
\end{acknowledgments}

\bibliography{ref}

\begin{thebibliography}{50}%
\makeatletter
\providecommand \@ifxundefined [1]{%
 \@ifx{#1\undefined}
}%
\providecommand \@ifnum [1]{%
 \ifnum #1\expandafter \@firstoftwo
 \else \expandafter \@secondoftwo
 \fi
}%
\providecommand \@ifx [1]{%
 \ifx #1\expandafter \@firstoftwo
 \else \expandafter \@secondoftwo
 \fi
}%
\providecommand \natexlab [1]{#1}%
\providecommand \enquote  [1]{``#1''}%
\providecommand \bibnamefont  [1]{#1}%
\providecommand \bibfnamefont [1]{#1}%
\providecommand \citenamefont [1]{#1}%
\providecommand \href@noop [0]{\@secondoftwo}%
\providecommand \href [0]{\begingroup \@sanitize@url \@href}%
\providecommand \@href[1]{\@@startlink{#1}\@@href}%
\providecommand \@@href[1]{\endgroup#1\@@endlink}%
\providecommand \@sanitize@url [0]{\catcode `\\12\catcode `\$12\catcode
  `\&12\catcode `\#12\catcode `\^12\catcode `\_12\catcode `\%12\relax}%
\providecommand \@@startlink[1]{}%
\providecommand \@@endlink[0]{}%
\providecommand \url  [0]{\begingroup\@sanitize@url \@url }%
\providecommand \@url [1]{\endgroup\@href {#1}{\urlprefix }}%
\providecommand \urlprefix  [0]{URL }%
\providecommand \Eprint [0]{\href }%
\providecommand \doibase [0]{https://doi.org/}%
\providecommand \selectlanguage [0]{\@gobble}%
\providecommand \bibinfo  [0]{\@secondoftwo}%
\providecommand \bibfield  [0]{\@secondoftwo}%
\providecommand \translation [1]{[#1]}%
\providecommand \BibitemOpen [0]{}%
\providecommand \bibitemStop [0]{}%
\providecommand \bibitemNoStop [0]{.\EOS\space}%
\providecommand \EOS [0]{\spacefactor3000\relax}%
\providecommand \BibitemShut  [1]{\csname bibitem#1\endcsname}%
\let\auto@bib@innerbib\@empty
\bibitem [{\citenamefont {Holstein}\ \emph {et~al.}(1973)\citenamefont
  {Holstein}, \citenamefont {Norton},\ and\ \citenamefont {Pincus}}]{holstein}%
  \BibitemOpen
  \bibfield  {author} {\bibinfo {author} {\bibfnamefont {T.}~\bibnamefont
  {Holstein}}, \bibinfo {author} {\bibfnamefont {R.~E.}\ \bibnamefont
  {Norton}},\ and\ \bibinfo {author} {\bibfnamefont {P.}~\bibnamefont
  {Pincus}},\ }\bibfield  {title} {\bibinfo {title} {de haas-van alphen effect
  and the specific heat of an electron gas},\ }\href
  {https://doi.org/10.1103/PhysRevB.8.2649} {\bibfield  {journal} {\bibinfo
  {journal} {Phys. Rev. B}\ }\textbf {\bibinfo {volume} {8}},\ \bibinfo {pages}
  {2649} (\bibinfo {year} {1973})}\BibitemShut {NoStop}%
\bibitem [{\citenamefont {Reizer}(1989)}]{reizer}%
  \BibitemOpen
  \bibfield  {author} {\bibinfo {author} {\bibfnamefont {M.~Y.}\ \bibnamefont
  {Reizer}},\ }\bibfield  {title} {\bibinfo {title} {Relativistic effects in
  the electron density of states, specific heat, and the electron spectrum of
  normal metals},\ }\href {https://doi.org/10.1103/PhysRevB.40.11571}
  {\bibfield  {journal} {\bibinfo  {journal} {Phys. Rev. B}\ }\textbf {\bibinfo
  {volume} {40}},\ \bibinfo {pages} {11571} (\bibinfo {year}
  {1989})}\BibitemShut {NoStop}%
\bibitem [{\citenamefont {Lee}\ and\ \citenamefont {Nagaosa}(1992)}]{leenag}%
  \BibitemOpen
  \bibfield  {author} {\bibinfo {author} {\bibfnamefont {P.~A.}\ \bibnamefont
  {Lee}}\ and\ \bibinfo {author} {\bibfnamefont {N.}~\bibnamefont {Nagaosa}},\
  }\bibfield  {title} {\bibinfo {title} {Gauge theory of the normal state of
  high-${\mathit{t}}_{\mathit{c}}$ superconductors},\ }\href
  {https://doi.org/10.1103/PhysRevB.46.5621} {\bibfield  {journal} {\bibinfo
  {journal} {Phys. Rev. B}\ }\textbf {\bibinfo {volume} {46}},\ \bibinfo
  {pages} {5621} (\bibinfo {year} {1992})}\BibitemShut {NoStop}%
\bibitem [{\citenamefont {Halperin}\ \emph {et~al.}(1993)\citenamefont
  {Halperin}, \citenamefont {Lee},\ and\ \citenamefont {Read}}]{HALPERIN}%
  \BibitemOpen
  \bibfield  {author} {\bibinfo {author} {\bibfnamefont {B.~I.}\ \bibnamefont
  {Halperin}}, \bibinfo {author} {\bibfnamefont {P.~A.}\ \bibnamefont {Lee}},\
  and\ \bibinfo {author} {\bibfnamefont {N.}~\bibnamefont {Read}},\ }\bibfield
  {title} {\bibinfo {title} {Theory of the half-filled landau level},\ }\href
  {https://doi.org/10.1103/PhysRevB.47.7312} {\bibfield  {journal} {\bibinfo
  {journal} {Phys. Rev. B}\ }\textbf {\bibinfo {volume} {47}},\ \bibinfo
  {pages} {7312} (\bibinfo {year} {1993})}\BibitemShut {NoStop}%
\bibitem [{\citenamefont {{Polchinski}}(1994)}]{polchinski}%
  \BibitemOpen
  \bibfield  {author} {\bibinfo {author} {\bibfnamefont {J.}~\bibnamefont
  {{Polchinski}}},\ }\bibfield  {title} {\bibinfo {title} {{Low-energy dynamics
  of the spinon-gauge system}},\ }\href
  {https://doi.org/10.1016/0550-3213(94)90449-9} {\bibfield  {journal}
  {\bibinfo  {journal} {Nuclear Physics B}\ }\textbf {\bibinfo {volume}
  {422}},\ \bibinfo {pages} {617} (\bibinfo {year} {1994})}\BibitemShut
  {NoStop}%
\bibitem [{\citenamefont {Altshuler}\ \emph {et~al.}(1994)\citenamefont
  {Altshuler}, \citenamefont {Ioffe},\ and\ \citenamefont
  {Millis}}]{ALTSHULER}%
  \BibitemOpen
  \bibfield  {author} {\bibinfo {author} {\bibfnamefont {B.~L.}\ \bibnamefont
  {Altshuler}}, \bibinfo {author} {\bibfnamefont {L.~B.}\ \bibnamefont
  {Ioffe}},\ and\ \bibinfo {author} {\bibfnamefont {A.~J.}\ \bibnamefont
  {Millis}},\ }\bibfield  {title} {\bibinfo {title} {Low-energy properties of
  fermions with singular interactions},\ }\href
  {https://doi.org/10.1103/PhysRevB.50.14048} {\bibfield  {journal} {\bibinfo
  {journal} {Phys. Rev. B}\ }\textbf {\bibinfo {volume} {50}},\ \bibinfo
  {pages} {14048} (\bibinfo {year} {1994})}\BibitemShut {NoStop}%
\bibitem [{\citenamefont {{Kim}}\ \emph {et~al.}(2008)\citenamefont {{Kim}},
  \citenamefont {{Lawler}}, \citenamefont {{Oreto}}, \citenamefont {{Sachdev}},
  \citenamefont {{Fradkin}},\ and\ \citenamefont {{Kivelson}}}]{YBKim}%
  \BibitemOpen
  \bibfield  {author} {\bibinfo {author} {\bibfnamefont {E.-A.}\ \bibnamefont
  {{Kim}}}, \bibinfo {author} {\bibfnamefont {M.~J.}\ \bibnamefont {{Lawler}}},
  \bibinfo {author} {\bibfnamefont {P.}~\bibnamefont {{Oreto}}}, \bibinfo
  {author} {\bibfnamefont {S.}~\bibnamefont {{Sachdev}}}, \bibinfo {author}
  {\bibfnamefont {E.}~\bibnamefont {{Fradkin}}},\ and\ \bibinfo {author}
  {\bibfnamefont {S.~A.}\ \bibnamefont {{Kivelson}}},\ }\bibfield  {title}
  {\bibinfo {title} {{Theory of the nodal nematic quantum phase transition in
  superconductors}},\ }\href {https://doi.org/10.1103/PhysRevB.77.184514}
  {\bibfield  {journal} {\bibinfo  {journal} {Phys. Rev. B}\ }\textbf {\bibinfo
  {volume} {77}},\ \bibinfo {eid} {184514} (\bibinfo {year}
  {2008})}\BibitemShut {NoStop}%
\bibitem [{\citenamefont {{Nayak}}\ and\ \citenamefont
  {{Wilczek}}(1994)}]{nayak}%
  \BibitemOpen
  \bibfield  {author} {\bibinfo {author} {\bibfnamefont {C.}~\bibnamefont
  {{Nayak}}}\ and\ \bibinfo {author} {\bibfnamefont {F.}~\bibnamefont
  {{Wilczek}}},\ }\bibfield  {title} {\bibinfo {title} {{Renormalization group
  approach to low temperature properties of a non-Fermi liquid metal}},\ }\href
  {https://doi.org/10.1016/0550-3213(94)90158-9} {\bibfield  {journal}
  {\bibinfo  {journal} {Nuclear Physics B}\ }\textbf {\bibinfo {volume}
  {430}},\ \bibinfo {pages} {534} (\bibinfo {year} {1994})}\BibitemShut
  {NoStop}%
\bibitem [{\citenamefont {{Lawler}}\ \emph {et~al.}(2006)\citenamefont
  {{Lawler}}, \citenamefont {{Barci}}, \citenamefont {{Fern{\'a}ndez}},
  \citenamefont {{Fradkin}},\ and\ \citenamefont {{Oxman}}}]{lawler1}%
  \BibitemOpen
  \bibfield  {author} {\bibinfo {author} {\bibfnamefont {M.~J.}\ \bibnamefont
  {{Lawler}}}, \bibinfo {author} {\bibfnamefont {D.~G.}\ \bibnamefont
  {{Barci}}}, \bibinfo {author} {\bibfnamefont {V.}~\bibnamefont
  {{Fern{\'a}ndez}}}, \bibinfo {author} {\bibfnamefont {E.}~\bibnamefont
  {{Fradkin}}},\ and\ \bibinfo {author} {\bibfnamefont {L.}~\bibnamefont
  {{Oxman}}},\ }\bibfield  {title} {\bibinfo {title} {{Nonperturbative behavior
  of the quantum phase transition to a nematic Fermi fluid}},\ }\href
  {https://doi.org/10.1103/PhysRevB.73.085101} {\bibfield  {journal} {\bibinfo
  {journal} {Phys. Rev. B}\ }\textbf {\bibinfo {volume} {73}},\ \bibinfo {eid}
  {085101} (\bibinfo {year} {2006})}\BibitemShut {NoStop}%
\bibitem [{\citenamefont {Lee}(2009)}]{SSLee}%
  \BibitemOpen
  \bibfield  {author} {\bibinfo {author} {\bibfnamefont {S.-S.}\ \bibnamefont
  {Lee}},\ }\bibfield  {title} {\bibinfo {title} {Low-energy effective theory
  of fermi surface coupled with u(1) gauge field in $2+1$ dimensions},\ }\href
  {https://doi.org/10.1103/PhysRevB.80.165102} {\bibfield  {journal} {\bibinfo
  {journal} {Phys. Rev. B}\ }\textbf {\bibinfo {volume} {80}},\ \bibinfo
  {pages} {165102} (\bibinfo {year} {2009})}\BibitemShut {NoStop}%
\bibitem [{\citenamefont {Metlitski}\ and\ \citenamefont
  {Sachdev}(2010{\natexlab{a}})}]{metlsach1}%
  \BibitemOpen
  \bibfield  {author} {\bibinfo {author} {\bibfnamefont {M.~A.}\ \bibnamefont
  {Metlitski}}\ and\ \bibinfo {author} {\bibfnamefont {S.}~\bibnamefont
  {Sachdev}},\ }\bibfield  {title} {\bibinfo {title} {Quantum phase transitions
  of metals in two spatial dimensions. i. ising-nematic order},\ }\href
  {https://doi.org/10.1103/PhysRevB.82.075127} {\bibfield  {journal} {\bibinfo
  {journal} {Phys. Rev. B}\ }\textbf {\bibinfo {volume} {82}},\ \bibinfo
  {pages} {075127} (\bibinfo {year} {2010}{\natexlab{a}})}\BibitemShut
  {NoStop}%
\bibitem [{\citenamefont {Metlitski}\ and\ \citenamefont
  {Sachdev}(2010{\natexlab{b}})}]{metlsach}%
  \BibitemOpen
  \bibfield  {author} {\bibinfo {author} {\bibfnamefont {M.~A.}\ \bibnamefont
  {Metlitski}}\ and\ \bibinfo {author} {\bibfnamefont {S.}~\bibnamefont
  {Sachdev}},\ }\bibfield  {title} {\bibinfo {title} {Quantum phase transitions
  of metals in two spatial dimensions. ii. spin density wave order},\ }\href
  {https://doi.org/10.1103/PhysRevB.82.075128} {\bibfield  {journal} {\bibinfo
  {journal} {Phys. Rev. B}\ }\textbf {\bibinfo {volume} {82}},\ \bibinfo
  {pages} {075128} (\bibinfo {year} {2010}{\natexlab{b}})}\BibitemShut
  {NoStop}%
\bibitem [{\citenamefont {{Abanov}}\ and\ \citenamefont
  {{Chubukov}}(2004)}]{chubukov1}%
  \BibitemOpen
  \bibfield  {author} {\bibinfo {author} {\bibfnamefont {A.}~\bibnamefont
  {{Abanov}}}\ and\ \bibinfo {author} {\bibfnamefont {A.}~\bibnamefont
  {{Chubukov}}},\ }\bibfield  {title} {\bibinfo {title} {{Anomalous Scaling at
  the Quantum Critical Point in Itinerant Antiferromagnets}},\ }\href
  {https://doi.org/10.1103/PhysRevLett.93.255702} {\bibfield  {journal}
  {\bibinfo  {journal} {Physical Review Letters}\ }\textbf {\bibinfo {volume}
  {93}},\ \bibinfo {eid} {255702} (\bibinfo {year} {2004})}\BibitemShut
  {NoStop}%
\bibitem [{\citenamefont {{Abanov}}\ and\ \citenamefont
  {{Chubukov}}(2000)}]{Chubukov}%
  \BibitemOpen
  \bibfield  {author} {\bibinfo {author} {\bibfnamefont {A.}~\bibnamefont
  {{Abanov}}}\ and\ \bibinfo {author} {\bibfnamefont {A.~V.}\ \bibnamefont
  {{Chubukov}}},\ }\bibfield  {title} {\bibinfo {title} {{Spin-Fermion Model
  near the Quantum Critical Point: One-Loop Renormalization Group Results}},\
  }\href {https://doi.org/10.1103/PhysRevLett.84.5608} {\bibfield  {journal}
  {\bibinfo  {journal} {Physical Review Letters}\ }\textbf {\bibinfo {volume}
  {84}},\ \bibinfo {pages} {5608} (\bibinfo {year} {2000})}\BibitemShut
  {NoStop}%
\bibitem [{\citenamefont {Mross}\ \emph {et~al.}(2010)\citenamefont {Mross},
  \citenamefont {McGreevy}, \citenamefont {Liu},\ and\ \citenamefont
  {Senthil}}]{mross}%
  \BibitemOpen
  \bibfield  {author} {\bibinfo {author} {\bibfnamefont {D.~F.}\ \bibnamefont
  {Mross}}, \bibinfo {author} {\bibfnamefont {J.}~\bibnamefont {McGreevy}},
  \bibinfo {author} {\bibfnamefont {H.}~\bibnamefont {Liu}},\ and\ \bibinfo
  {author} {\bibfnamefont {T.}~\bibnamefont {Senthil}},\ }\bibfield  {title}
  {\bibinfo {title} {Controlled expansion for certain non-fermi-liquid
  metals},\ }\href {https://doi.org/10.1103/PhysRevB.82.045121} {\bibfield
  {journal} {\bibinfo  {journal} {Phys. Rev. B}\ }\textbf {\bibinfo {volume}
  {82}},\ \bibinfo {pages} {045121} (\bibinfo {year} {2010})}\BibitemShut
  {NoStop}%
\bibitem [{\citenamefont {{Jiang}}\ \emph {et~al.}(2013)\citenamefont
  {{Jiang}}, \citenamefont {{Block}}, \citenamefont {{Mishmash}}, \citenamefont
  {{Garrison}}, \citenamefont {{Sheng}}, \citenamefont {{Motrunich}},\ and\
  \citenamefont {{Fisher}}}]{Jiang}%
  \BibitemOpen
  \bibfield  {author} {\bibinfo {author} {\bibfnamefont {H.-C.}\ \bibnamefont
  {{Jiang}}}, \bibinfo {author} {\bibfnamefont {M.~S.}\ \bibnamefont
  {{Block}}}, \bibinfo {author} {\bibfnamefont {R.~V.}\ \bibnamefont
  {{Mishmash}}}, \bibinfo {author} {\bibfnamefont {J.~R.}\ \bibnamefont
  {{Garrison}}}, \bibinfo {author} {\bibfnamefont {D.~N.}\ \bibnamefont
  {{Sheng}}}, \bibinfo {author} {\bibfnamefont {O.~I.}\ \bibnamefont
  {{Motrunich}}},\ and\ \bibinfo {author} {\bibfnamefont {M.~P.~A.}\
  \bibnamefont {{Fisher}}},\ }\bibfield  {title} {\bibinfo {title}
  {{Non-Fermi-liquid d-wave metal phase of strongly interacting electrons}},\
  }\href {https://doi.org/10.1038/nature11732} {\bibfield  {journal} {\bibinfo
  {journal} {\nat}\ }\textbf {\bibinfo {volume} {493}},\ \bibinfo {pages} {39}
  (\bibinfo {year} {2013})}\BibitemShut {NoStop}%
\bibitem [{\citenamefont {Sur}\ and\ \citenamefont {Lee}(2014)}]{Shouvik1}%
  \BibitemOpen
  \bibfield  {author} {\bibinfo {author} {\bibfnamefont {S.}~\bibnamefont
  {Sur}}\ and\ \bibinfo {author} {\bibfnamefont {S.-S.}\ \bibnamefont {Lee}},\
  }\bibfield  {title} {\bibinfo {title} {Chiral non-fermi liquids},\ }\href
  {https://doi.org/10.1103/PhysRevB.90.045121} {\bibfield  {journal} {\bibinfo
  {journal} {Phys. Rev. B}\ }\textbf {\bibinfo {volume} {90}},\ \bibinfo
  {pages} {045121} (\bibinfo {year} {2014})}\BibitemShut {NoStop}%
\bibitem [{\citenamefont {Dalidovich}\ and\ \citenamefont
  {Lee}(2013)}]{Lee-Dalid}%
  \BibitemOpen
  \bibfield  {author} {\bibinfo {author} {\bibfnamefont {D.}~\bibnamefont
  {Dalidovich}}\ and\ \bibinfo {author} {\bibfnamefont {S.-S.}\ \bibnamefont
  {Lee}},\ }\bibfield  {title} {\bibinfo {title} {Perturbative non-fermi
  liquids from dimensional regularization},\ }\href
  {https://doi.org/10.1103/PhysRevB.88.245106} {\bibfield  {journal} {\bibinfo
  {journal} {Phys. Rev. B}\ }\textbf {\bibinfo {volume} {88}},\ \bibinfo
  {pages} {245106} (\bibinfo {year} {2013})}\BibitemShut {NoStop}%
\bibitem [{\citenamefont {Sur}\ and\ \citenamefont {Lee}(2015)}]{shouvik2}%
  \BibitemOpen
  \bibfield  {author} {\bibinfo {author} {\bibfnamefont {S.}~\bibnamefont
  {Sur}}\ and\ \bibinfo {author} {\bibfnamefont {S.-S.}\ \bibnamefont {Lee}},\
  }\bibfield  {title} {\bibinfo {title} {Quasilocal strange metal},\ }\href
  {https://doi.org/10.1103/PhysRevB.91.125136} {\bibfield  {journal} {\bibinfo
  {journal} {Phys. Rev. B}\ }\textbf {\bibinfo {volume} {91}},\ \bibinfo
  {pages} {125136} (\bibinfo {year} {2015})}\BibitemShut {NoStop}%
\bibitem [{\citenamefont {Oganesyan}\ \emph {et~al.}(2001)\citenamefont
  {Oganesyan}, \citenamefont {Kivelson},\ and\ \citenamefont
  {Fradkin}}]{ogankivfr}%
  \BibitemOpen
  \bibfield  {author} {\bibinfo {author} {\bibfnamefont {V.}~\bibnamefont
  {Oganesyan}}, \bibinfo {author} {\bibfnamefont {S.~A.}\ \bibnamefont
  {Kivelson}},\ and\ \bibinfo {author} {\bibfnamefont {E.}~\bibnamefont
  {Fradkin}},\ }\bibfield  {title} {\bibinfo {title} {Quantum theory of a
  nematic fermi fluid},\ }\href {https://doi.org/10.1103/PhysRevB.64.195109}
  {\bibfield  {journal} {\bibinfo  {journal} {Phys. Rev. B}\ }\textbf {\bibinfo
  {volume} {64}},\ \bibinfo {pages} {195109} (\bibinfo {year}
  {2001})}\BibitemShut {NoStop}%
\bibitem [{\citenamefont {Metzner}\ \emph {et~al.}(2003)\citenamefont
  {Metzner}, \citenamefont {Rohe},\ and\ \citenamefont
  {Andergassen}}]{metzner}%
  \BibitemOpen
  \bibfield  {author} {\bibinfo {author} {\bibfnamefont {W.}~\bibnamefont
  {Metzner}}, \bibinfo {author} {\bibfnamefont {D.}~\bibnamefont {Rohe}},\ and\
  \bibinfo {author} {\bibfnamefont {S.}~\bibnamefont {Andergassen}},\
  }\bibfield  {title} {\bibinfo {title} {Soft fermi surfaces and breakdown of
  fermi-liquid behavior},\ }\href
  {https://doi.org/10.1103/PhysRevLett.91.066402} {\bibfield  {journal}
  {\bibinfo  {journal} {Phys. Rev. Lett.}\ }\textbf {\bibinfo {volume} {91}},\
  \bibinfo {pages} {066402} (\bibinfo {year} {2003})}\BibitemShut {NoStop}%
\bibitem [{\citenamefont {Dell'Anna}\ and\ \citenamefont
  {Metzner}(2006)}]{delanna}%
  \BibitemOpen
  \bibfield  {author} {\bibinfo {author} {\bibfnamefont {L.}~\bibnamefont
  {Dell'Anna}}\ and\ \bibinfo {author} {\bibfnamefont {W.}~\bibnamefont
  {Metzner}},\ }\bibfield  {title} {\bibinfo {title} {Fermi surface
  fluctuations and single electron excitations near pomeranchuk instability in
  two dimensions},\ }\href {https://doi.org/10.1103/PhysRevB.73.045127}
  {\bibfield  {journal} {\bibinfo  {journal} {Phys. Rev. B}\ }\textbf {\bibinfo
  {volume} {73}},\ \bibinfo {pages} {045127} (\bibinfo {year}
  {2006})}\BibitemShut {NoStop}%
\bibitem [{\citenamefont {Kee}\ \emph {et~al.}(2003)\citenamefont {Kee},
  \citenamefont {Kim},\ and\ \citenamefont {Chung}}]{kee}%
  \BibitemOpen
  \bibfield  {author} {\bibinfo {author} {\bibfnamefont {H.-Y.}\ \bibnamefont
  {Kee}}, \bibinfo {author} {\bibfnamefont {E.~H.}\ \bibnamefont {Kim}},\ and\
  \bibinfo {author} {\bibfnamefont {C.-H.}\ \bibnamefont {Chung}},\ }\bibfield
  {title} {\bibinfo {title} {Signatures of an electronic nematic phase at the
  isotropic-nematic phase transition},\ }\href
  {https://doi.org/10.1103/PhysRevB.68.245109} {\bibfield  {journal} {\bibinfo
  {journal} {Phys. Rev. B}\ }\textbf {\bibinfo {volume} {68}},\ \bibinfo
  {pages} {245109} (\bibinfo {year} {2003})}\BibitemShut {NoStop}%
\bibitem [{\citenamefont {Rech}\ \emph {et~al.}(2006)\citenamefont {Rech},
  \citenamefont {P\'epin},\ and\ \citenamefont {Chubukov}}]{rech}%
  \BibitemOpen
  \bibfield  {author} {\bibinfo {author} {\bibfnamefont {J.}~\bibnamefont
  {Rech}}, \bibinfo {author} {\bibfnamefont {C.}~\bibnamefont {P\'epin}},\ and\
  \bibinfo {author} {\bibfnamefont {A.~V.}\ \bibnamefont {Chubukov}},\
  }\bibfield  {title} {\bibinfo {title} {Quantum critical behavior in itinerant
  electron systems: Eliashberg theory and instability of a ferromagnetic
  quantum critical point},\ }\href {https://doi.org/10.1103/PhysRevB.74.195126}
  {\bibfield  {journal} {\bibinfo  {journal} {Phys. Rev. B}\ }\textbf {\bibinfo
  {volume} {74}},\ \bibinfo {pages} {195126} (\bibinfo {year}
  {2006})}\BibitemShut {NoStop}%
\bibitem [{\citenamefont {{W{\"o}lfle}}\ and\ \citenamefont
  {{Rosch}}(2007)}]{wolfle}%
  \BibitemOpen
  \bibfield  {author} {\bibinfo {author} {\bibfnamefont {P.}~\bibnamefont
  {{W{\"o}lfle}}}\ and\ \bibinfo {author} {\bibfnamefont {A.}~\bibnamefont
  {{Rosch}}},\ }\bibfield  {title} {\bibinfo {title} {{Fermi Liquid Near a
  Quantum Critical Point}},\ }\href {https://doi.org/10.1007/s10909-007-9308-y}
  {\bibfield  {journal} {\bibinfo  {journal} {Journal of Low Temperature
  Physics}\ }\textbf {\bibinfo {volume} {147}},\ \bibinfo {pages} {165}
  (\bibinfo {year} {2007})}\BibitemShut {NoStop}%
\bibitem [{\citenamefont {Maslov}\ and\ \citenamefont
  {Chubukov}(2010)}]{maslov}%
  \BibitemOpen
  \bibfield  {author} {\bibinfo {author} {\bibfnamefont {D.~L.}\ \bibnamefont
  {Maslov}}\ and\ \bibinfo {author} {\bibfnamefont {A.~V.}\ \bibnamefont
  {Chubukov}},\ }\bibfield  {title} {\bibinfo {title} {Fermi liquid near
  pomeranchuk quantum criticality},\ }\href
  {https://doi.org/10.1103/PhysRevB.81.045110} {\bibfield  {journal} {\bibinfo
  {journal} {Phys. Rev. B}\ }\textbf {\bibinfo {volume} {81}},\ \bibinfo
  {pages} {045110} (\bibinfo {year} {2010})}\BibitemShut {NoStop}%
\bibitem [{\citenamefont {{Quintanilla}}\ and\ \citenamefont
  {{Schofield}}(2006)}]{quintanilla}%
  \BibitemOpen
  \bibfield  {author} {\bibinfo {author} {\bibfnamefont {J.}~\bibnamefont
  {{Quintanilla}}}\ and\ \bibinfo {author} {\bibfnamefont {A.~J.}\ \bibnamefont
  {{Schofield}}},\ }\bibfield  {title} {\bibinfo {title} {{Pomeranchuk and
  topological Fermi surface instabilities from central interactions}},\ }\href
  {https://doi.org/10.1103/PhysRevB.74.115126} {\bibfield  {journal} {\bibinfo
  {journal} {Phys. Rev. B}\ }\textbf {\bibinfo {volume} {74}},\ \bibinfo {eid}
  {115126} (\bibinfo {year} {2006})}\BibitemShut {NoStop}%
\bibitem [{\citenamefont {{Yamase}}\ and\ \citenamefont
  {{Kohno}}(2000)}]{yamase1}%
  \BibitemOpen
  \bibfield  {author} {\bibinfo {author} {\bibfnamefont {H.}~\bibnamefont
  {{Yamase}}}\ and\ \bibinfo {author} {\bibfnamefont {H.}~\bibnamefont
  {{Kohno}}},\ }\bibfield  {title} {\bibinfo {title} {{Instability toward
  Formation of Quasi-One-Dimensional Fermi Surface in Two-Dimensional t-J
  Model}},\ }\href {https://doi.org/10.1143/JPSJ.69.2151} {\bibfield  {journal}
  {\bibinfo  {journal} {Journal of the Physical Society of Japan}\ }\textbf
  {\bibinfo {volume} {69}},\ \bibinfo {pages} {2151} (\bibinfo {year}
  {2000})}\BibitemShut {NoStop}%
\bibitem [{\citenamefont {Yamase}\ \emph {et~al.}(2005)\citenamefont {Yamase},
  \citenamefont {Oganesyan},\ and\ \citenamefont {Metzner}}]{yamase2}%
  \BibitemOpen
  \bibfield  {author} {\bibinfo {author} {\bibfnamefont {H.}~\bibnamefont
  {Yamase}}, \bibinfo {author} {\bibfnamefont {V.}~\bibnamefont {Oganesyan}},\
  and\ \bibinfo {author} {\bibfnamefont {W.}~\bibnamefont {Metzner}},\
  }\bibfield  {title} {\bibinfo {title} {Mean-field theory for
  symmetry-breaking fermi surface deformations on a square lattice},\ }\href
  {https://doi.org/10.1103/PhysRevB.72.035114} {\bibfield  {journal} {\bibinfo
  {journal} {Phys. Rev. B}\ }\textbf {\bibinfo {volume} {72}},\ \bibinfo
  {pages} {035114} (\bibinfo {year} {2005})}\BibitemShut {NoStop}%
\bibitem [{\citenamefont {Halboth}\ and\ \citenamefont
  {Metzner}(2000)}]{halboth}%
  \BibitemOpen
  \bibfield  {author} {\bibinfo {author} {\bibfnamefont {C.~J.}\ \bibnamefont
  {Halboth}}\ and\ \bibinfo {author} {\bibfnamefont {W.}~\bibnamefont
  {Metzner}},\ }\bibfield  {title} {\bibinfo {title} {$\mathit{d}$-wave
  superconductivity and pomeranchuk instability in the two-dimensional hubbard
  model},\ }\href {https://doi.org/10.1103/PhysRevLett.85.5162} {\bibfield
  {journal} {\bibinfo  {journal} {Phys. Rev. Lett.}\ }\textbf {\bibinfo
  {volume} {85}},\ \bibinfo {pages} {5162} (\bibinfo {year}
  {2000})}\BibitemShut {NoStop}%
\bibitem [{\citenamefont {Jakubczyk}\ \emph {et~al.}(2008)\citenamefont
  {Jakubczyk}, \citenamefont {Strack}, \citenamefont {Katanin},\ and\
  \citenamefont {Metzner}}]{jakub}%
  \BibitemOpen
  \bibfield  {author} {\bibinfo {author} {\bibfnamefont {P.}~\bibnamefont
  {Jakubczyk}}, \bibinfo {author} {\bibfnamefont {P.}~\bibnamefont {Strack}},
  \bibinfo {author} {\bibfnamefont {A.~A.}\ \bibnamefont {Katanin}},\ and\
  \bibinfo {author} {\bibfnamefont {W.}~\bibnamefont {Metzner}},\ }\bibfield
  {title} {\bibinfo {title} {Renormalization group for phases with broken
  discrete symmetry near quantum critical points},\ }\href
  {https://doi.org/10.1103/PhysRevB.77.195120} {\bibfield  {journal} {\bibinfo
  {journal} {Phys. Rev. B}\ }\textbf {\bibinfo {volume} {77}},\ \bibinfo
  {pages} {195120} (\bibinfo {year} {2008})}\BibitemShut {NoStop}%
\bibitem [{\citenamefont {Zacharias}\ \emph {et~al.}(2009)\citenamefont
  {Zacharias}, \citenamefont {W\"olfle},\ and\ \citenamefont
  {Garst}}]{zacharias}%
  \BibitemOpen
  \bibfield  {author} {\bibinfo {author} {\bibfnamefont {M.}~\bibnamefont
  {Zacharias}}, \bibinfo {author} {\bibfnamefont {P.}~\bibnamefont
  {W\"olfle}},\ and\ \bibinfo {author} {\bibfnamefont {M.}~\bibnamefont
  {Garst}},\ }\bibfield  {title} {\bibinfo {title} {Multiscale quantum
  criticality: Pomeranchuk instability in isotropic metals},\ }\href
  {https://doi.org/10.1103/PhysRevB.80.165116} {\bibfield  {journal} {\bibinfo
  {journal} {Phys. Rev. B}\ }\textbf {\bibinfo {volume} {80}},\ \bibinfo
  {pages} {165116} (\bibinfo {year} {2009})}\BibitemShut {NoStop}%
\bibitem [{\citenamefont {Kim}\ \emph {et~al.}(2008)\citenamefont {Kim},
  \citenamefont {Lawler}, \citenamefont {Oreto}, \citenamefont {Sachdev},
  \citenamefont {Fradkin},\ and\ \citenamefont {Kivelson}}]{kim}%
  \BibitemOpen
  \bibfield  {author} {\bibinfo {author} {\bibfnamefont {E.-A.}\ \bibnamefont
  {Kim}}, \bibinfo {author} {\bibfnamefont {M.~J.}\ \bibnamefont {Lawler}},
  \bibinfo {author} {\bibfnamefont {P.}~\bibnamefont {Oreto}}, \bibinfo
  {author} {\bibfnamefont {S.}~\bibnamefont {Sachdev}}, \bibinfo {author}
  {\bibfnamefont {E.}~\bibnamefont {Fradkin}},\ and\ \bibinfo {author}
  {\bibfnamefont {S.~A.}\ \bibnamefont {Kivelson}},\ }\bibfield  {title}
  {\bibinfo {title} {Theory of the nodal nematic quantum phase transition in
  superconductors},\ }\href {https://doi.org/10.1103/PhysRevB.77.184514}
  {\bibfield  {journal} {\bibinfo  {journal} {Phys. Rev. B}\ }\textbf {\bibinfo
  {volume} {77}},\ \bibinfo {pages} {184514} (\bibinfo {year}
  {2008})}\BibitemShut {NoStop}%
\bibitem [{\citenamefont {{Huh}}\ and\ \citenamefont {{Sachdev}}(2008)}]{huh}%
  \BibitemOpen
  \bibfield  {author} {\bibinfo {author} {\bibfnamefont {Y.}~\bibnamefont
  {{Huh}}}\ and\ \bibinfo {author} {\bibfnamefont {S.}~\bibnamefont
  {{Sachdev}}},\ }\bibfield  {title} {\bibinfo {title} {{Renormalization group
  theory of nematic ordering in d -wave superconductors}},\ }\href
  {https://doi.org/10.1103/PhysRevB.78.064512} {\bibfield  {journal} {\bibinfo
  {journal} {Phys. Rev. B}\ }\textbf {\bibinfo {volume} {78}},\ \bibinfo {eid}
  {064512} (\bibinfo {year} {2008})}\BibitemShut {NoStop}%
\bibitem [{\citenamefont {{Motrunich}}(2005)}]{MOTRUNICH}%
  \BibitemOpen
  \bibfield  {author} {\bibinfo {author} {\bibfnamefont {O.~I.}\ \bibnamefont
  {{Motrunich}}},\ }\bibfield  {title} {\bibinfo {title} {{Variational study of
  triangular lattice spin- 1/2 model with ring exchanges and spin liquid state
  in {$\kappa$}- (ET)$_{2}$ Cu$_{2}$ (CN)$_{3}$}},\ }\href
  {https://doi.org/10.1103/PhysRevB.72.045105} {\bibfield  {journal} {\bibinfo
  {journal} {Phys. Rev. B}\ }\textbf {\bibinfo {volume} {72}},\ \bibinfo {eid}
  {045105} (\bibinfo {year} {2005})}\BibitemShut {NoStop}%
\bibitem [{\citenamefont {Lee}\ and\ \citenamefont {Lee}(2005)}]{LEE_U1}%
  \BibitemOpen
  \bibfield  {author} {\bibinfo {author} {\bibfnamefont {S.-S.}\ \bibnamefont
  {Lee}}\ and\ \bibinfo {author} {\bibfnamefont {P.~A.}\ \bibnamefont {Lee}},\
  }\bibfield  {title} {\bibinfo {title} {U(1) gauge theory of the hubbard
  model: Spin liquid states and possible application to
  $\ensuremath{\kappa}\mathrm{\text{-}}(\mathrm{BEDT}\mathrm{\text{-}}\mathrm{TTF}{)}_{2}{\mathrm{cu}}_{2}(\mathrm{CN}{)}_{3}$},\
  }\href {https://doi.org/10.1103/PhysRevLett.95.036403} {\bibfield  {journal}
  {\bibinfo  {journal} {Phys. Rev. Lett.}\ }\textbf {\bibinfo {volume} {95}},\
  \bibinfo {pages} {036403} (\bibinfo {year} {2005})}\BibitemShut {NoStop}%
\bibitem [{\citenamefont {{Lee}}\ \emph {et~al.}(2006)\citenamefont {{Lee}},
  \citenamefont {{Nagaosa}},\ and\ \citenamefont {{Wen}}}]{PALEE}%
  \BibitemOpen
  \bibfield  {author} {\bibinfo {author} {\bibfnamefont {P.~A.}\ \bibnamefont
  {{Lee}}}, \bibinfo {author} {\bibfnamefont {N.}~\bibnamefont {{Nagaosa}}},\
  and\ \bibinfo {author} {\bibfnamefont {X.-G.}\ \bibnamefont {{Wen}}},\
  }\bibfield  {title} {\bibinfo {title} {{Doping a Mott insulator: Physics of
  high-temperature superconductivity}},\ }\href
  {https://doi.org/10.1103/RevModPhys.78.17} {\bibfield  {journal} {\bibinfo
  {journal} {Reviews of Modern Physics}\ }\textbf {\bibinfo {volume} {78}},\
  \bibinfo {pages} {17} (\bibinfo {year} {2006})}\BibitemShut {NoStop}%
\bibitem [{\citenamefont {Motrunich}\ and\ \citenamefont
  {Fisher}(2007)}]{MotrunichFisher}%
  \BibitemOpen
  \bibfield  {author} {\bibinfo {author} {\bibfnamefont {O.~I.}\ \bibnamefont
  {Motrunich}}\ and\ \bibinfo {author} {\bibfnamefont {M.~P.~A.}\ \bibnamefont
  {Fisher}},\ }\bibfield  {title} {\bibinfo {title} {$d$-wave correlated
  critical bose liquids in two dimensions},\ }\href
  {https://doi.org/10.1103/PhysRevB.75.235116} {\bibfield  {journal} {\bibinfo
  {journal} {Phys. Rev. B}\ }\textbf {\bibinfo {volume} {75}},\ \bibinfo
  {pages} {235116} (\bibinfo {year} {2007})}\BibitemShut {NoStop}%
\bibitem [{\citenamefont {Senthil}(2008)}]{senthil}%
  \BibitemOpen
  \bibfield  {author} {\bibinfo {author} {\bibfnamefont {T.}~\bibnamefont
  {Senthil}},\ }\bibfield  {title} {\bibinfo {title} {Critical fermi surfaces
  and non-fermi liquid metals},\ }\href
  {https://doi.org/10.1103/PhysRevB.78.035103} {\bibfield  {journal} {\bibinfo
  {journal} {Phys. Rev. B}\ }\textbf {\bibinfo {volume} {78}},\ \bibinfo
  {pages} {035103} (\bibinfo {year} {2008})}\BibitemShut {NoStop}%
\bibitem [{\citenamefont {{Lee}}(2008)}]{LEE2008}%
  \BibitemOpen
  \bibfield  {author} {\bibinfo {author} {\bibfnamefont {S.-S.}\ \bibnamefont
  {{Lee}}},\ }\bibfield  {title} {\bibinfo {title} {{Stability of the U(1) spin
  liquid with a spinon Fermi surface in 2+1 dimensions}},\ }\href
  {https://doi.org/10.1103/PhysRevB.78.085129} {\bibfield  {journal} {\bibinfo
  {journal} {Phys. Rev. B}\ }\textbf {\bibinfo {volume} {78}},\ \bibinfo {eid}
  {085129} (\bibinfo {year} {2008})}\BibitemShut {NoStop}%
\bibitem [{Note1()}]{Note1}%
  \BibitemOpen
  \bibinfo {note} {If there is a superconducting instability, the locality in
  momentum space breaks down even for $m=1$. Here we assume $d-m>1$, for which
  there is no perturbative pairing instability.}\BibitemShut {Stop}%
\bibitem [{\citenamefont {Fitzpatrick}\ \emph {et~al.}(2013)\citenamefont
  {Fitzpatrick}, \citenamefont {Kachru}, \citenamefont {Kaplan},\ and\
  \citenamefont {Raghu}}]{Fit}%
  \BibitemOpen
  \bibfield  {author} {\bibinfo {author} {\bibfnamefont {A.~L.}\ \bibnamefont
  {Fitzpatrick}}, \bibinfo {author} {\bibfnamefont {S.}~\bibnamefont {Kachru}},
  \bibinfo {author} {\bibfnamefont {J.}~\bibnamefont {Kaplan}},\ and\ \bibinfo
  {author} {\bibfnamefont {S.}~\bibnamefont {Raghu}},\ }\bibfield  {title}
  {\bibinfo {title} {Non-fermi-liquid fixed point in a wilsonian theory of
  quantum critical metals},\ }\href
  {https://doi.org/10.1103/PhysRevB.88.125116} {\bibfield  {journal} {\bibinfo
  {journal} {Phys. Rev. B}\ }\textbf {\bibinfo {volume} {88}},\ \bibinfo
  {pages} {125116} (\bibinfo {year} {2013})}\BibitemShut {NoStop}%
\bibitem [{\citenamefont {Torroba}\ and\ \citenamefont {Wang}(2014)}]{Torroba}%
  \BibitemOpen
  \bibfield  {author} {\bibinfo {author} {\bibfnamefont {G.}~\bibnamefont
  {Torroba}}\ and\ \bibinfo {author} {\bibfnamefont {H.}~\bibnamefont {Wang}},\
  }\bibfield  {title} {\bibinfo {title} {Quantum critical metals in
  $4\ensuremath{-}\ensuremath{\epsilon}$ dimensions},\ }\href
  {https://doi.org/10.1103/PhysRevB.90.165144} {\bibfield  {journal} {\bibinfo
  {journal} {Phys. Rev. B}\ }\textbf {\bibinfo {volume} {90}},\ \bibinfo
  {pages} {165144} (\bibinfo {year} {2014})}\BibitemShut {NoStop}%
\bibitem [{\citenamefont {{Chakravarty}}\ \emph {et~al.}(1995)\citenamefont
  {{Chakravarty}}, \citenamefont {{Norton}},\ and\ \citenamefont
  {{Sylju{\aa}sen}}}]{Chakravarty}%
  \BibitemOpen
  \bibfield  {author} {\bibinfo {author} {\bibfnamefont {S.}~\bibnamefont
  {{Chakravarty}}}, \bibinfo {author} {\bibfnamefont {R.~E.}\ \bibnamefont
  {{Norton}}},\ and\ \bibinfo {author} {\bibfnamefont {O.~F.}\ \bibnamefont
  {{Sylju{\aa}sen}}},\ }\bibfield  {title} {\bibinfo {title} {{Transverse Gauge
  Interactions and the Vanquished Fermi Liquid}},\ }\href
  {https://doi.org/10.1103/PhysRevLett.74.1423} {\bibfield  {journal} {\bibinfo
   {journal} {Physical Review Letters}\ }\textbf {\bibinfo {volume} {74}},\
  \bibinfo {pages} {1423} (\bibinfo {year} {1995})}\BibitemShut {NoStop}%
\bibitem [{\citenamefont {Senthil}\ and\ \citenamefont
  {Shankar}(2009)}]{senshank}%
  \BibitemOpen
  \bibfield  {author} {\bibinfo {author} {\bibfnamefont {T.}~\bibnamefont
  {Senthil}}\ and\ \bibinfo {author} {\bibfnamefont {R.}~\bibnamefont
  {Shankar}},\ }\bibfield  {title} {\bibinfo {title} {Fermi surfaces in general
  codimension and a new controlled nontrivial fixed point},\ }\href
  {https://doi.org/10.1103/PhysRevLett.102.046406} {\bibfield  {journal}
  {\bibinfo  {journal} {Phys. Rev. Lett.}\ }\textbf {\bibinfo {volume} {102}},\
  \bibinfo {pages} {046406} (\bibinfo {year} {2009})}\BibitemShut {NoStop}%
\bibitem [{\citenamefont {Sch\"afer}\ and\ \citenamefont
  {Schwenzer}(2004)}]{schafer}%
  \BibitemOpen
  \bibfield  {author} {\bibinfo {author} {\bibfnamefont {T.}~\bibnamefont
  {Sch\"afer}}\ and\ \bibinfo {author} {\bibfnamefont {K.}~\bibnamefont
  {Schwenzer}},\ }\bibfield  {title} {\bibinfo {title} {Non-fermi liquid
  effects in qcd at high density},\ }\href
  {https://doi.org/10.1103/PhysRevD.70.054007} {\bibfield  {journal} {\bibinfo
  {journal} {Phys. Rev. D}\ }\textbf {\bibinfo {volume} {70}},\ \bibinfo
  {pages} {054007} (\bibinfo {year} {2004})}\BibitemShut {NoStop}%
\bibitem [{\citenamefont {Chung}\ \emph {et~al.}(2013)\citenamefont {Chung},
  \citenamefont {Mandal}, \citenamefont {Raghu},\ and\ \citenamefont
  {Chakravarty}}]{ips-sudip1}%
  \BibitemOpen
  \bibfield  {author} {\bibinfo {author} {\bibfnamefont {S.~B.}\ \bibnamefont
  {Chung}}, \bibinfo {author} {\bibfnamefont {I.}~\bibnamefont {Mandal}},
  \bibinfo {author} {\bibfnamefont {S.}~\bibnamefont {Raghu}},\ and\ \bibinfo
  {author} {\bibfnamefont {S.}~\bibnamefont {Chakravarty}},\ }\bibfield
  {title} {\bibinfo {title} {Higher angular momentum pairing from transverse
  gauge interactions},\ }\href {https://doi.org/10.1103/PhysRevB.88.045127}
  {\bibfield  {journal} {\bibinfo  {journal} {Phys. Rev. B}\ }\textbf {\bibinfo
  {volume} {88}},\ \bibinfo {pages} {045127} (\bibinfo {year}
  {2013})}\BibitemShut {NoStop}%
\bibitem [{\citenamefont {Wang}\ \emph {et~al.}(2014)\citenamefont {Wang},
  \citenamefont {Mandal}, \citenamefont {Chung},\ and\ \citenamefont
  {Chakravarty}}]{ips-sudip2}%
  \BibitemOpen
  \bibfield  {author} {\bibinfo {author} {\bibfnamefont {Z.}~\bibnamefont
  {Wang}}, \bibinfo {author} {\bibfnamefont {I.}~\bibnamefont {Mandal}},
  \bibinfo {author} {\bibfnamefont {S.~B.}\ \bibnamefont {Chung}},\ and\
  \bibinfo {author} {\bibfnamefont {S.}~\bibnamefont {Chakravarty}},\
  }\bibfield  {title} {\bibinfo {title} {Pairing in half-filled landau level},\
  }\href {https://doi.org/http://dx.doi.org/10.1016/j.aop.2014.09.021}
  {\bibfield  {journal} {\bibinfo  {journal} {Annals of Physics}\ }\textbf
  {\bibinfo {volume} {351}},\ \bibinfo {pages} {727 } (\bibinfo {year}
  {2014})}\BibitemShut {NoStop}%
\bibitem [{\citenamefont {Metlitski}\ \emph {et~al.}(2015)\citenamefont
  {Metlitski}, \citenamefont {Mross}, \citenamefont {Sachdev},\ and\
  \citenamefont {Senthil}}]{Max}%
  \BibitemOpen
  \bibfield  {author} {\bibinfo {author} {\bibfnamefont {M.~A.}\ \bibnamefont
  {Metlitski}}, \bibinfo {author} {\bibfnamefont {D.~F.}\ \bibnamefont
  {Mross}}, \bibinfo {author} {\bibfnamefont {S.}~\bibnamefont {Sachdev}},\
  and\ \bibinfo {author} {\bibfnamefont {T.}~\bibnamefont {Senthil}},\
  }\bibfield  {title} {\bibinfo {title} {Cooper pairing in non-fermi liquids},\
  }\href {https://doi.org/10.1103/PhysRevB.91.115111} {\bibfield  {journal}
  {\bibinfo  {journal} {Phys. Rev. B}\ }\textbf {\bibinfo {volume} {91}},\
  \bibinfo {pages} {115111} (\bibinfo {year} {2015})}\BibitemShut {NoStop}%
\bibitem [{\citenamefont {Mandal}(2016)}]{ips-sc}%
  \BibitemOpen
  \bibfield  {author} {\bibinfo {author} {\bibfnamefont {I.}~\bibnamefont
  {Mandal}},\ }\bibfield  {title} {\bibinfo {title} {Superconducting
  instability in non-fermi liquids},\ }\href
  {https://doi.org/10.1103/PhysRevB.94.115138} {\bibfield  {journal} {\bibinfo
  {journal} {Phys. Rev. B}\ }\textbf {\bibinfo {volume} {94}},\ \bibinfo
  {pages} {115138} (\bibinfo {year} {2016})}\BibitemShut {NoStop}%
\end{thebibliography}%

\begin{widetext}
\appendix

\section{Computation of the Feynman Diagrams at One-loop}
\label{app:oneloop}

\subsection{One-loop boson self-energy}

In this section, we compute the one-loop boson self-energy :
\bqa 
\label{bosloop0}
\Pi_1 (q) &=& -(ie)^2 \mu^x
\int dk \, \mbox{Tr}
\left[ \gamma_{d-m} G_0 (k+q)\gamma_{d-m} G_0 (k) \right] \,,
\eqa
where the bare fermion propagator is given by 
$G_0 (k) =\frac{1}{i} \, \frac{\vec \Gamma \cdot \vec K +
\gamma_{d-m} \delta_k} 
{\vec K^2  + \delta_k^2} 
\exp \Big \lbrace - \frac {{\vec{L}}_{(k)}^2}  { \mu \, {\tilde{k}}_F } \Big \rbrace$.
Performing the integration over $k_{d-m}$, we obtain
\begin{eqnarray}
\Pi_1 (q) 
&=&  e^2 \, \mu^x \int \frac{ d{\vec{L}}_{(k)} \, d\vec K}{(2\pi)^d} 
\frac{ \big(\,|\vec K +\vec Q|+|\vec K| \, \big)\,  
\left[ \,\vec K \cdot (\vec K +\vec Q) - |\vec K|\, |\vec K +\vec Q| \, \right]  \,
\exp \left ({-\frac{{\vec{L}}_{(k)}^2 + {\vec{L}}_{(k+q)}^2} { \mu \, {\tilde{k}}_F }} \right )
}
{  |\vec K|\,|\vec K +\vec Q|\, 
\left[ \big (\, \delta_q +2 \, q_{d-m+1} \, k_{d-m+1} \, \big )^2 + \big(\, |\vec K +\vec Q|+|\vec K|\, \big)^2 
\right] }    \,, \nonumber
\end{eqnarray}
where we have chosen a coordinate system such that ${\vec{L}}_{(q)} = (q_{d-m+1},0, 0,\ldots,0) $, without loss of generality. 
Because of the rotational symmetry in  ${\vec L}_{(q)}$,
$\Pi_1(q)$ depends only on its magnitude.
From the expression for the integration over $k_{d-m+1}$,
\bqa
\label{I}
{I} &\equiv&
\int \frac{d k_{d-m+1}}{2 \pi} 
\frac{   \exp \left [  {-\frac{2 \,  k_{d-m+1}  \, \left (  \, k_{d-m+1}  \,+ \,  |\vec{L}_{(q)}| \, \right) + |\vec{L}_{(q)}| ^2 } 
 {k_F}}   
\right ] } 
{ \left (\, \delta_q +2 \,  k_{d-m+1} \,  |\vec{L}_{(q)}| \, \right )^2 + \big(\, |\vec K +\vec Q|+|\vec K|\, \big)^2 }  \nn
&=& 
\exp \left ({- \frac{ |\vec{L}_{(q)}| ^2} {2 k_F} } \right)
\frac { 1 } 
{ |\vec{L}_{(q)}|^2  \sqrt{8 \, k_F}  } 
\times
 \mathcal{F} \left(  \frac{ |\vec K +\vec Q|+|\vec K|}
{ |\vec{L}_{(q)}|   \sqrt{2 \, k_F } } ,
\frac{q_{d-m}} { |\vec{L}_{(q)}|   \sqrt{2 \, k_F } }
\right)
\,,
\eqa
with
\beq
\label{F}
\mathcal{F} ( y , u) =
\int_{-\infty}^{\infty}
\frac{d z} {2\pi}  \,
\frac{  
\exp \left ({- z^2 } \right)
}
{ \left ( z + u  \right )^2 + y^2 } \, ,
\eeq
and 
$\label{J}
J \equiv \int_{-\infty}^{\infty} \frac{dz }{2 \pi} \, \exp{ \left ( - \frac{2z ^2} { \mu \, {\tilde{k}}_F } \right) } =\sqrt{\frac{ \mu \, {\tilde{k}}_F }{8 \pi}}$
for the integration over the remaining components of ${\vec L}_{(k)}$,
the self-energy is written as:
\begin{eqnarray}
\Pi_1 (q) 
&=& \frac{e^2 \, \mu^x J^{m-1}
\, \exp \left ({- \frac{ |\vec{L}_{(q)}| ^2} {2 k_F} } \right) } 
 {|\vec{L}_{(q)}|^2  \sqrt{8 \, k_F} \,  (2\pi)^{d-m} } 
  \int  d\vec K
\left\{
\frac{\vec K \cdot (\vec K +\vec Q)}{|\vec K|\,\,|\vec K +\vec Q|}
-1 \right\}
\left (\,|\vec K +\vec Q|+|\vec K| \, \right )
 \, \mathcal{F} \left(  \frac{ |\vec K +\vec Q|+|\vec K|}
{ |\vec{L}_{(q)}|   \sqrt{2 \, k_F } } ,
\frac{q_{d-m}} { |\vec{L}_{(q)}|   \sqrt{2 \, k_F } }
\right).\nn
\end{eqnarray}

It is difficult to obtain the exact expression of $\Pi_1(q)$
for general values of $k_F$ and $q$.
Here we focus on the limit of physical importance
where $k_F$ is much larger than all other scales, including external momentum and the UV cut-off of $\vec K$.
In this limit, we use
$\mathcal{F} (y, u) 
 \simeq \frac{1} {2 \,|y|}$
for $y, u \ll 1$ to simplify the expression 
for the self-energy to
\beq\label{pia11}
\Pi_1 (q) =\frac{   e^2 \, \mu^x } {  2^{m+1} \, |\vec{L}_{(q)}|}  \Big ( \frac{ \mu \, {\tilde{k}}_F } {2 \pi} \Big )^{\frac{m-1}{2}} \, \, \, I_1 (d-m, \vec Q)  \,, 
\eeq
where
\beq
I_1 (d-m, \vec Q) = \int \frac{d\vec K}{(2\pi)^{d-m}} 
\left\{
\frac{\vec K \cdot (\vec K +\vec Q)}{|\vec K|\,\,|\vec K +\vec Q|}
-1 \right\}. 
\label{a8}
\eeq
Using the Feynman parametrization
\beq\label{feynm}
\frac{1}{A^{n_1} B^{n_2 }}= 
\frac{\Gamma (n_1 + n_2)}  {\Gamma (n_1 ) \, \Gamma ( n_2)}
\int_0^1 \frac{t^{ n_1 -1}\,(1-t )^{ n_2-1}\,dt}
{\left[ \, t \, A +(1-t ) \, B \,  \right]^{ n_1 + n_2}} \,,
\eeq
with $n_1 = n_2 =1/2$,
$A=| \vec K + \vec Q |^2$ and
$B=| \vec K |^2$
we rewrite \eq{a8} as
\beq
I_1 (d-m, \vec Q) = \frac{1}
{ \pi \, (2 \pi)^{d-m}  } 
\int_0^1 \frac{dt \ }{\sqrt{ t \, (1- t )}}\,
\int {d\vec K} \,
\left\{
\frac{\vec K \cdot (\vec K+\vec Q)}
{  t \, |\vec K  + \vec Q|^2 +  (1- t ) \, \vec K^2 } 
-1 \right\}. \nonumber
\eeq
The integrations over $\vec K$ and $t$ give
\begin{eqnarray}
I_1(d-m, \vec Q) 
&=&  - \frac{\Gamma^2(\frac{d-m+1}{2}) \, |\vec Q|^{d-m}  }  {2^{d-m-1}  \, \pi^{\frac{d-m} {2} } \, \Gamma(\frac{d-m}{2})   \, \Gamma(d-m+1) \, |\cos \lbrace  \frac{\pi (d-m+1)} {2} \rbrace|  } \, , 
\end{eqnarray}
for $0 < d-m < 2$,
and  the self-energy is obtained to be 
\beq
\Pi_1 (q) =- \beta_d \, e^2 \, \mu^x \,  \frac{|\vec Q|^{d-m} \, ( \mu \, {\tilde{k}}_F )^{ \frac{m-1}{2} }}{|\vec{L}_{(q)}|} \, ,
\label{api}
\eeq
where $\beta_d$ is defined after Eq.~(\ref{babos}). 

The expression in Eq.~(\ref{api}) is valid for 
 $\frac{|\vec Q|}{|\vec{L}_{(q)} | \, \sqrt{  k_F}} \ll 1$
and $  \frac{q_{d-m}} { |\vec{L}_{(q)}|   \sqrt{  k_F } } \ll 1$. 
On the other hand, it is no longer valid 
in a slim region of $q$ satisfying
 $\frac{|\vec Q|}{|\vec{L}_{(q)} | \, \sqrt{  k_F}} \gg 1$
or $  \frac{q_{d-m}} { |\vec{L}_{(q)}|   \sqrt{  k_F } } \gg 1$,
when $k_F$ is large but finite.
Now let us consider the boson self-energy in this region.
When $\frac{|\vec Q|}{|\vec{L}_{(q)} | \, \sqrt{  k_F}} \gg 1$
or $  \frac{q_{d-m}} { |\vec{L}_{(q)}|   \sqrt{  k_F } } \gg 1$,
we need to use the expression
\beq
\mathcal{F} (y, u) \simeq \frac{1}{y^2 + u^2} \int_{-\infty}^{\infty}
\frac{dz} {2 \pi} \exp{(-z^2)}
=\frac{1}{ 2  \sqrt{ \pi} \,( y^2 +u^2)} 
\eeq
for $u \gg 1$ or $y \gg 1$,
and the self-energy becomes 
\bqa
\Pi_1 (q) &=&
\frac{  e^2 \, \mu^x \, J^{m-1} \sqrt{ k_F} }{  {(2\pi)^{d-m}}  \sqrt{8 \pi} }
\left ( t_1 + t_2 \right ) \,,
\nn
\eqa
where 
\bqa
t_1 &=& \int_{  |\vec K|< |\vec Q|}
{d\vec K} \, \,  \,
T \left( \vec K ,\vec Q, q_{d-m} \right) \,,\\
t_2 &=& \int_{  |\vec K|> |\vec Q|}
{d\vec K} \, \, \,
 T \left( \vec K ,\vec Q, q_{d-m} \right) \,, 
\eqa
with $T \left( \vec K ,\vec Q, q_{d-m} \right) = \left\{
\frac{\vec K \cdot (\vec K +\vec Q)}{|\vec K|\,\,|\vec K +\vec Q|}
-1 \right\}
\frac{ |\vec K +\vec Q|+|\vec K| }
{ \left( \,  |\vec K +\vec Q|+|\vec K| \, \right )^2   +   q_{d-m}^2 }$.
We can estimate $t_1$ and $t_2$ as
\bqa
t_1  \sim  - \frac{ |\vec Q|^{d-m+1} } { |\vec Q|^2  +  q_{d-m}^2 } \,, 
~~~~~
 t_2  \sim  - |\vec Q|^2  \, |q_{d-m}|^{d-m-3} \,\, g\left( \frac{|\vec Q|}{|q_{d-m}|} \right)\,, 
\eqa
where $g(t) = \int_t^{\infty} dv \, \frac{v^{d-m-2}} { 1+ \, 4 v^2}$.
Hence, near $|\vec{L}_{(q)} | =0$, the boson self-energy becomes
\beq
\label{pi2}
\Pi_1    (q) \simeq - e^2 \, \mu^x \, J^{m-1} \sqrt{ k_F} \,
\tilde f( |\vec Q|, q_{d-m}) \,,
\eeq
where 
\beq
\tilde f( |\vec Q|, q_{d-m}) \sim
\left[
\frac{ |\vec Q|^{d-m+1} } {  |\vec Q|^2  +  q_{d-m}^2 }
+\tilde C \, |\vec Q|^2  \, |q_{d-m}|^{d-m-3} \, \, g\left( \frac{|\vec Q|}{|q_{d-m}|} \right)
\right ],
\eeq
with $ \tilde C > 0$, a constant.
The full one-loop boson self-energy crosses over 
from \eq{api} to \eq{pi2} around
$\frac{|\vec Q|}{|\vec{L}_{(q)} | \, \sqrt{  k_F}}, \frac{q_{d-m}} { |\vec{L}_{(q)}|   \sqrt{  k_F } } \sim 1$.
However, the range of $q$ in which \eq{pi2} is needed 
becomes vanishingly small in the large $k_F$ limit.
As a result, we can use \eq{api} 
for the dressed boson propagator 
to the leading order in $1/k_F$
as will be shown explicitly in the next section.



\subsection{One-loop fermion self-energy}
\label{app:oneloopferm}

Here we compute the one-loop fermion self-energy:
\begin{eqnarray}
\label{sigma1ex}
\Sigma_1 (q) &=& \frac{(ie)^2  \, \mu^{x}}  {N} 
\int dk \,
\gamma_{d-m} \, G_0 (q-k)\, \gamma_{d-m} \,D_1 (k) \,.
\label{sigma1}
\end{eqnarray} 
We will first compute \eq{sigma1} using
$D_1(k)$ dressed with the boson-self energy in Eq.~(\ref{api}).
Integrating over 
$k_{d-m}\,,$ we obtain
 \beq
\Sigma_1 (q) = \frac{i \,e^2 \,  \mu^{x}}{ 2 N}  \int \frac{d \vec{K} }{(2\pi)^{d-m}} \, 
\frac{ \vec \Gamma \cdot (\vec K -\vec Q)}
{|\vec K -\vec Q|}   \, \times {I}_2 \, , \nonumber
\eeq
where
\bqa
I_2 
 &=& \int \frac{d \Omega_m} {(2 \pi)^{m-1}}  
\int_0^{\infty} \frac{d{ |\vec{L}_{(k)}| }} {2 \pi} \,
 \frac {  |\vec{L}_{(k)}|^{m-1} }    
 {  |\vec{L}_{(k)}|^2  - \Pi_1  (k)  } \nn
&=&  \frac{  (\pi)^{\frac{2-m}{2}} }
{3 \times 2^{m-1} \, \Gamma{(m/2)} \,  | \sin \lbrace (m+1) \pi/3 \rbrace | \, \lbrace  \, \beta_{d} \, e^2 \, \mu^{x} \,( \mu \, {\tilde{k}}_F )^{ \frac{m-1}{2}}  \,  |\vec K|^{d-m} \, \rbrace ^{\frac{2-m}{3}} } \,.
\label{I2}
\end{eqnarray}
Here we have assumed that $0 < m+1 < 3$, for which the $|\vec{L}_{(k)}|$ integral converges 
without the exponential damping factor 
in the fermion propagator.
Therefore, we obtain
\beq\label{ferm11}
\Sigma_1 (q) = \frac{i \, e^{2(m+1)/3} \, \mu^{ x \, (m+1)/3} \,   \pi^{(2-m)/2} \times I_3 (d-m, \vec Q)  }
{6 N 
\times 2^{m-1} \, \Gamma{(m/2)} \,  | \sin \lbrace (m+1) \pi/3 \rbrace |\,  \beta_{d}^{(2-m)/3}  \, ( \mu \, {\tilde{k}}_F ) ^{(m-1)(2-m)/6} } 
\eeq
to the leading order in $1/k_F$, where
\beq
\label{i3}
I_3 (d-m, \vec Q) = \int \frac{d\vec K}{(2\pi)^{d-m}}
\frac{\vec \Gamma\cdot (\vec K -\vec Q)}{|\vec K|^{(d-m)(2-m)/3}\,|\vec K -\vec Q|} \,.
\eeq
Using the Feynman parametrization (\ref{feynm}),
we obtain 
\bqa
\label{intII}
I_3 (d-m, \vec Q) &=& \frac{\Gamma ( \frac{1}{2} + \beta)}
{\Gamma (\beta) \sqrt{\pi} \, (2 \pi)^{d-m}  } 
\int_0^1 \frac{dt \,(1-t )^{\beta-1} }{\sqrt{ t }}\,
\int {d\vec K} \,
\frac{\vec \Gamma \cdot (\vec K- \vec Q)}
{   \big[ \, t \, |\vec K  -\vec Q|^2 +  (1- t ) \vec K^2 \, \big ]^{ \frac {1}{2} +  \beta} } \nn
&=& - \frac{ \Gamma ( \beta - \frac{d-m-1}{2}) \, \Gamma (  \frac{d-m-2 \beta}{2}) \,  \Gamma (  \frac{d-m+1}{2}) \, \vec \Gamma \cdot \vec Q }
{\Gamma (\beta) \, \Gamma{(d-m-\beta + \frac{1}{2}}) \,  \sqrt{\pi} \, (4 \pi)^{\frac{d-m}{2} }  \, (\vec Q ^2)^{ \beta - \frac{d-m-1 }{2}}  }  \,,
\eqa
where  $\beta \equiv \frac{(d-m)(2-m)} {6}$.
%
Finally, the fermion self-energy is obtained to be
\begin{eqnarray} 
\label{one-loop-ferm}
\Sigma_1 (q) &=& - \frac{ i \, e^{2(m+1)/3} \, \mu^{ x \, (m+1)/3} \, \vec \Gamma \cdot \vec Q    }
{6 N \,   \pi^{(m-1)/2}
(4 \pi)^{\frac{d-m}{2}} \,2^{m-1} \,  | \sin \lbrace (m+1) \pi/3 \rbrace |\,  \beta_{d}^{(2-m)/3}  \,( \mu \, {\tilde{k}}_F ) ^{(m-1)(2-m)/6} \, (\vec Q ^2)^{  \frac{3-  (m+1)(d-m) }{6}}}  \nn
&& \times \frac{  \Gamma ( \frac{3-  (m+1)(d-m) }{6}) \, \Gamma (  \frac{d-m-2 \beta}{2}) \,  \Gamma (  \frac{d-m+1}{2})  } 
{  \Gamma{(m/2)} \, \Gamma(\beta) \,  \Gamma ( d-m -  \beta + \frac{1}{2}) } \,.
\end{eqnarray}

The above expression has been obtained by using 
the boson self-energy in \eq{api}, which is valid in the large $k_F$ limit.
Now we explicitly check that our use of \eq{api} in \eq{sigma1}
is valid to the leading order in $1/k_F$.
Suppose we use the exact expression of the one-loop boson self-energy, which is valid for all $k_F$, to compute the fermion self-energy:
\bqa
\Sigma_1^{exact} (q) &=& \frac{i \,e^2 \,  \mu^{x}}{N}  
\int dk \,
 \frac{ \exp \left(-\frac{{\vec{L}}_{(q-k)}^2} {k_F} \right) } { {\vec{L}}_{(k)}^2  - \Pi_1 ^{exact}   (k) }
 \frac{\gamma_{d-m} \, \delta_{q-k} -\vec \Gamma \cdot (\vec Q -\vec K)}
{(\vec Q -\vec K)^2 +\delta_{q-k}^2}  \,.
\eqa
$\Pi_1^{exact}$ deviates from $\Pi_1$ in \eq{api} 
for $|{\vec{L}}_{(k)}| < \delta \equiv \frac{\Lambda}{\sqrt{k_F}} \ll \sqrt{\Lambda} \ll \sqrt{k_F}$,
where $\Lambda$ is the UV cut-off for $\vec K$.
In this region, we can safely ignore the exponential damping factor and the contribution of $|{\vec{L}}_{(k)}|$ in $\delta_{q-k}$.
In this case, we have
\bqa
\Sigma_1^{exact} (q) - \Sigma_1(q) &\simeq &
 \frac{i \,e^2 \,  \mu^{x}}{N}
\int \frac{d \vec{K}  \, d k_{d-m}} { (2 \pi)^{d+1}  } 
\int_{| {\vec{L}}_{(k)}|=0}^{| {\vec{L}}_{(k)}| = \delta}
d {\vec{L}}_{(k)}  \,
 \frac{\gamma_{d-m} \, \delta_{q-k} -\vec \Gamma \cdot (\vec Q -\vec K)}
{(\vec Q -\vec K)^2 +\delta_{q-k}^2} 
\left [
\frac{ 1}
 {  {\vec{L}}_{(k)}^2  - \Pi_1^{exact} (k)} 
-
\frac{ 1}
 { {\vec{L}}_{(k)}^2  - \Pi_1  (k)  }
\right] \nn
&\simeq &
 \frac{i \,e^2 \,  \mu^{x}}{N}    \int d\Omega_{m} 
\int \frac{d \vec{K}  \, d k_{d-m}} { (2 \pi)^{d+1}  } \,
\left [
 \frac{\gamma_{d-m} \, \delta_{q-k}-\vec \Gamma \cdot (\vec Q -\vec K)}
{(\vec Q -\vec K)^2 +\delta_{q-k}^2} 
\right ]  \Bigg|_{ {\vec{L}}_{(k)}=0 } 
\left ( \,  i_{21} -i_{22}  \, \right ) \,,
\eqa
where 
\beq
i_{21}  = \int_{| {\vec{L}}_{(k)}|=0}^{| {\vec{L}}_{(k)}| = \delta}
 \frac{ d |{\vec{L}}_{(k)}| \, \, |{\vec{L}}_{(k)}|^{m-1}  }
 { 
 {\vec{L}}_{(k)}^2  + e^2 \, \mu^x \, J^{m-1} \sqrt{ k_F} \, \,
\tilde{f} \left( |\vec K |, k_{d-m} \right)
 }  \,, \quad
 i_{22} =
  \int_{| {\vec{L}}_{(k)}|=0}^{| {\vec{L}}_{(k)}| = \delta }
 \frac{  d  |{\vec{L}}_{(k)}| \, \, |{\vec{L}}_{(k)}|^{m-1}}
 { {\vec{L}}_{(k)}^2  - \Pi_1  (k)  } \, .
\eeq
Integrating over $|{\vec L}_{(k)}|$, we obtain
 \begin{eqnarray}
\label{i21}
 i_{21}
&=&  \, _2F_1 \left( 
1,\frac{m}{2} ;\frac{m+2}{2}; 
- \frac{\delta^2}
{ e^2 \, \mu^x \, J^{m-1} \sqrt{ k_F} \, \,
\tilde{f} \left( |\vec K |, k_{d-m} \right)}
\right)
\,
 \frac{\delta^m}
{ m \, e^2 \, \mu^x \, J^{m-1} \sqrt{ k_F} \, \,
\tilde{f} \left( |\vec K |, k_{d-m} \right)} \, ,\\
\label{i22}
 i_{22} &=&   _2F_1 \left( 
1,\frac{m+1}{3} ;\frac{m+4}{3}; 
- \frac{\delta^3}
{ \beta_{d} \, e^2 \, \mu^{ x} \, k_F^{  \frac{m-1}{2} }  \, |\vec{K}|^{ (d-m)}   }\,
\right)
\,
\frac{\delta^{m+1}}
{ (m+1) \, \beta_{d} \, e^2 \, \mu^{x} \,k_F^{ \frac{m-1}{2} } \, |\vec{K}|^{(d-m)}  } \,,
\end{eqnarray}
where $_2F_1$ is the hypergeometric function.
In the limit ${k_F} \rightarrow \infty$ with fixed $\Lambda$,
we have $\frac{i_{21}}{I_2}, \frac{i_{22}}{I_2} \sim k_F^{-\frac{m^2-m+2}{2}} \rightarrow 0$,
where $I_2$ is defined in \eq{I2}.
Therefore, $\Sigma^{exact} (q) $ goes to $ \Sigma (q) $ in the large $k_F$ limit,
and Eq.~(\ref{one-loop-ferm}) is valid to the leading order in $1/k_F$.

\section{Computation of the Feynman Diagrams at Two-loop}
\label{app:twoloop}

\begin{figure}[ht]
\begin{center}
\includegraphics[scale=0.5]{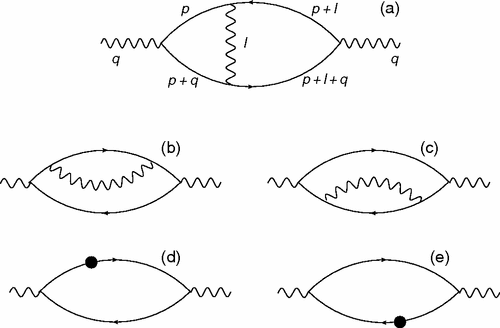}
\end{center}
\caption{The diagrams for two-loop boson self-energy.}
\label{fig:bos2}
\end{figure}

\begin{figure}[ht]
\begin{center}
\includegraphics[scale=0.5]{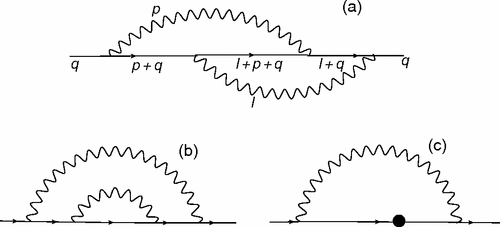}
\end{center}
\caption{The diagrams for two-loop fermion self-energy.}
\label{fig:ferm2}
\end{figure}

\begin{figure}[h!]
\begin{center}
\includegraphics[scale=0.55]{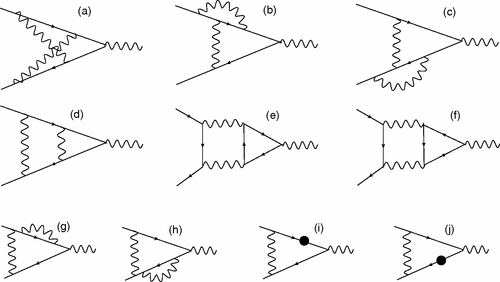}
\end{center}
\caption{The diagrams for two-loop vertex corrections.}
\label{fig:vert2}
\end{figure} 

All the two-loop diagrams are shown in 
Figs.~\ref{fig:bos2},\ref{fig:ferm2} and \ref{fig:vert2}.  
The black circles in  
Figs.~\ref{fig:bos2} (d)-(e), \ref{fig:ferm2}(c) and 
\ref{fig:vert2}(i)-(j) denote the one-loop
counterterm for the fermion self-energy,
\beq
\label{2legv}
i \, A_1^{(1)} \bar \Psi (\vec \Gamma \cdot \vec Q ) \Psi.
\eeq 
Among the self-energy diagrams, only Figs.~\ref{fig:bos2}(a) and \ref{fig:ferm2}(a) contribute\cite{Lee-Dalid}.
The vertex correction can be obtained from the  fermion self-energy correction through the Ward identity.
As we will see through explicit computations in the following sections,
the two-loop diagrams are suppressed not only by $\tilde e$ but also by $\Lambda/k_F$ in the low-energy limit for $m>1$.


\subsection{Two-loop contribution to boson self-energy}
\label{2loopbos}

We compute the two-loop boson self-energy shown 
in Fig.~\ref{fig:bos2} (a):
\beq
\label{prbos2a}
\Pi_2 (q) = -\frac{e^4 \mu^{2\, x }  \, N } {N^2}
 \int {dl\, dp} 
\, D_1 (l) \, {\rm Tr} \{ \gamma_{d-m}\, G_0 (p)\, \gamma_{d-m}\, G_0 (p+l)\, 
\gamma_{d-m} \,G_0 (p+l+q) \,\gamma_{d-m} \,G_0 (p+q) \}\,.
\eeq
Taking the trace, we obtain
\beq\label{prbos2b}
\Pi_2 (q) = -\frac{e^4 \mu^{2\, x}}{N}  \,
\int {dl\, dp} \,
D_1 (l)\,
\frac{{\cal B}_{1}}{{\cal D}_{1}} \,
\exp\left( {-\frac{{\vec{L}}_{(p)}^2 + {\vec{L}}_{(p+q)}^2 +  {\vec{L}}_{(p+l)}^2+  {\vec{L}}_{(p+l+q)}^2} { k_F }} \right)\,,
\eeq
where
\bqa
{\cal B}_{1}& =& 2 \,[ \,\delta_{p+l} \, \delta_{p+q+l} -(\vec P +\vec L)
\cdot (\vec P +\vec L +\vec Q)\,]\, 
 \, [ \, \delta_{p+q} \, \delta_{p} -(\vec P +\vec Q)
\cdot \vec P \,] - 2 \, [\, (\vec P +\vec L)\cdot (\vec P +\vec Q)\, ]\,
[ \,(\vec P +\vec L +\vec Q)\cdot \vec P] 
\nonumber \\
&& 
+ 2 \, [ \,(\vec P +\vec L)\cdot \vec P] \,
[ \,(\vec P +\vec L +\vec Q)\cdot (\vec P +\vec Q)\, ] 
- 2 \, [\, \delta_{p+l} \,  (\vec P +\vec L +\vec Q) + \delta_{p+l+q}  \, (\vec P +\vec L)\, ]
\cdot [\, \delta_{p+q} \, \vec P + \delta_{p} \, (\vec P +\vec Q)\, ] \,, 
\label{B1} \nonumber \\
{\cal D}_{1} &=& [\, \delta_{p}^2 +\vec P^2] \,  [ \, \delta_{p+q}^2 +(\vec P +\vec Q)^2\,] \,
[ \, \delta_{p+l}^2 +(\vec P +\vec L)^2]\,
[ \, \delta_{p+l+q}^2 +(\vec P +\vec L +\vec Q)^2]\,. \label{D1}
\eqa
Shifting the variables as
\beq
p_{d-m } \rightarrow p_{d-m } - \vec L_{(p)}^2\,, \qquad 
l_{d-m } \rightarrow l_{d-m } - p_{d-m } -  \vec L_{(p+l)}^2\,,\nonumber
\eeq
we can substitute
\beq
\delta_{p} \rightarrow p_{d-m} \,, \quad \delta_{p+q} \rightarrow p_{d-m}
+2 \,  \vec L_{(p)} \cdot  \vec L_{(q)} +\delta_q \,, \quad
\delta_{l+p} \rightarrow l_{d-m } \,, \quad \delta_{p+l+q} \rightarrow l_{d-m }
+2 \, \vec L_{(p+l)} \cdot  \vec L_{(q)}  +\delta_q \,. \nonumber
\eeq
Integration over $p_{d-m }$ and $l_{d-m }$ gives us
\beq
\Pi_2 (q) = -\frac{e^4 \mu^{2\, x }}{N} 
\int \frac{d \vec L_{(l)} \, d\vec L\, d\vec L_{(p)}\, d\vec P } {(2\pi)^{2\,d}}
\, D_1 (l)\,
\frac{{\cal B}_{2}}{{\cal D}_{2}}\,
\exp\left( {-\frac{{\vec{L}}_{(p)}^2 + {\vec{L}}_{(p+q)}^2 +  {\vec{L}}_{(p+l)}^2+  {\vec{L}}_{(p+l+q)}^2} { k_F }} \right)\,,
\eeq
where
\bqa
{\cal B}_{2} &=& 2\, \left(\, |\vec P +\vec L|+ |\vec P +\vec L +\vec Q|\, \right)
\left( \,|\vec P +\vec Q|+ |\vec P|\, \right)\nonumber\\
&& \quad \times\, \left. \Bigl\{ 
\left[ \,|\vec P +\vec L| \,|\vec P +\vec L +\vec Q|-
(\vec P +\vec L) \cdot (\vec P +\vec L +\vec Q)\, \right]
\left[ \,|\vec P +\vec Q|\, |\vec P|-
(\vec P +\vec Q) \cdot \vec P  \,  \right] \right.\nonumber\\
&&  \quad \quad \quad - \left. [\, (\vec P +\vec L)\cdot (\vec P +\vec Q)\, ]\,
[ \, (\vec P +\vec L+\vec Q)\cdot \vec P \,]
+ [\, (\vec P +\vec L)\cdot \vec P\, ]\,
[ \, (\vec P +\vec L+\vec Q)\cdot (\vec P +\vec Q)\, ] \right. \Bigr\} \nonumber\\
&&  - 2\, \left( \,2 \,  \vec L_{(p+l)} \cdot  \vec L_{(q)}+\delta_q \right) 
\left( \, 2 \,  \vec L_{(p)} \cdot  \vec L_{(q)} +\delta_q \,\right) \nn
&&  \quad \times\,   \left[\, |\vec P +\vec L +\vec Q |\, (\vec P +\vec L)- 
|\vec P +\vec L| (\vec P +\vec L +\vec Q )\right]\cdot
\left[ \,|\vec P +\vec Q |\, \vec P - 
|\vec P|\, (\vec P +\vec Q )\,\right], \label{B2} \\
{\cal D}_{2} &=& 4 \,  | \vec P|\,|\vec P +\vec Q |
  \,|\vec P +\vec L| \, |\vec P +\vec Q +\vec L| \,
 \left[ \left( \, 2 \, \vec L_{(p+l)} \cdot  \vec L_{(q)}  +\delta_q \,\right)^2 + 
\left(|\vec P +\vec L|+ |\vec P +\vec Q +\vec L| \right)^2  \,
\right] \nonumber\\
&&  \times \, \left[ \, \left(\, 2 \, \vec L_{(p)} \cdot  \vec L_{(q)}  +\delta_q \right)^2 + 
\left(|\vec P |+ |\vec P +\vec Q | \right)^2  \right]. \label{D2}
\eqa
Without loss of generality, we can choose the coordinate system such that ${\vec{L}}_{(q)} = ( q_{d-m+1},0, 0,\ldots,0) $ with $q_{d-m+1}>0$.
After making a further change of variables as
\bqa
\vec L  \rightarrow   \vec L -\vec P, \quad
\vec P \rightarrow   \vec P -\frac{\vec Q}{2}\,, \quad
2 \, | \vec L_{(q)} |\, p_{d-m+1} + \, \delta_q   \rightarrow  p_{d-m+1}  \,,
\eqa
and integrating over $p_{d-m+1} $ (neglecting the corresponding exponential damping part), we obtain:
\bqa
\label{prbos2d}
\Pi_2 (q) &\simeq&  -\frac{e^4 \mu^{2\, x}}{N}  \,
\int \frac{d \vec L_{(l)} \, d\vec L}{(2\pi)^{d}} \frac{{d \vec{u}}_{(p)} \, d\vec P}{(2\pi)^{d-1}} \,
D_1(   \vec L_{(l)} ,|\vec L -\vec P| )\,
\frac{{\cal B}_{3}(\vec L, \vec P, \vec Q)}
{{\cal D}_{3} (l, \vec P, q)} 
\, \exp\left( {-\frac{   3\, {\vec{u}}_{(p)}^2 } { k_F }} \right)\nn
&\simeq &-\frac{e^4 \mu^{2\, x}}{N}  \,\left(\frac{k_F}{12 \, \pi} \right)^{\frac {m-1} {2}}
\int \frac{d \vec L_{(l)} \, d\vec L}{(2\pi)^{d}} \frac{ d\vec P}{(2\pi)^{d-m}} \,
D_1(   \vec L_{(l)} ,|\vec L -\vec P| )\,
\frac{{\cal B}_{3}(\vec L, \vec P, \vec Q)}
{{\cal D}_{3} (l, \vec P, q)} \,,\nn
\eqa
where
\bqa
 {\vec{u}}_{(k)} &=&(k_{d-m+2},\ldots,k_d)\,,\nn
{\cal B}_{3} (\vec L, \vec P, \vec Q) &=&
{\cal B}_{4} (\vec L, \vec P, \vec Q)
\,  \bar {\cal D } (\vec L, \vec P, \vec Q)\,,
 \label{B3} \\
{\cal D}_{3} (l, \vec P, q) & = &
8\,| \vec L_{(q)} | \,
{\cal D}_{4} (\vec L, \vec P, \vec Q)
\,  \Bigl\{ \bar {\cal D }^2 (\vec L, \vec P, \vec Q) + 4  ( \vec L_{(q)} \cdot \vec L_{(l)}\,)^2  \Bigr\} \,,
\label{D3}  \nn
{\cal B}_{4} (\vec L, \vec P, \vec Q) &=&
 \left( \,|\vec L -\vec Q/2|\,|\vec L +\vec Q/2| 
-\vec L^2 +\vec Q^2/4 \, \right)
\left( \,|\vec P -\vec Q/2| \, |\vec P +\vec Q/2| 
-\vec P^2 +\vec Q^2/4 \, \right) \nonumber\\
&& \quad -   \left. \left[ \, (\vec L -\vec Q/2)\cdot (\vec P +\vec Q/2)\, \right]
\left[\,(\vec L +\vec Q/2)\cdot (\vec P -\vec Q/2)\, \right]\right.\nonumber\\ 
&& \quad +    \left. \left[\, (\vec L -\vec Q/2)\cdot (\vec P -\vec Q/2)\, \right]
\left[ \,(\vec L +\vec Q/2)\cdot (\vec P +\vec Q/2)\, \right]
\right. \nonumber\\
&&  \quad  -  \left. 
 |\vec L +\vec Q/2|\, |\vec P +\vec Q/2|\,
[ \,(\vec L -\vec Q/2 )\cdot (\vec P -\vec Q/2 )\, ] \right.\nonumber\\
&&  \quad +  \left. |\vec L +\vec Q/2|\, |\vec P -\vec Q/2|\,
[\, (\vec L -\vec Q/2 )\cdot (\vec P +\vec Q/2 )\, ] \right.\nonumber\\
&&  \quad  + \left. |\vec L -\vec Q/2|\, |\vec P +\vec Q/2|\,
[ (\vec L +\vec Q/2 )\cdot (\vec P -\vec Q/2 )\,]\right.\nonumber\\
&&   \quad - |\vec L -\vec Q/2|\, |\vec P -\vec Q/2|\,
[ \,(\vec L +\vec Q/2 )\cdot (\vec P +\vec Q/2 )\,] \,, \label{B4} \\
{\cal D}_{4} (\vec L, \vec P, \vec Q) & = &
 |\vec L-\vec Q/2|\, |\vec L +\vec Q/2| \,
|\vec P-\vec Q/2|\, |\vec P +\vec Q/2 |\,, 
\label{D4} \\
\bar {\cal D } (\vec L, \vec P, \vec Q)  &=&
|\vec L-\vec Q/2|+ |\vec L+\vec Q/2| + 
|\vec P-\vec Q/2|+ |\vec P+\vec Q/2| \,.
\eqa
Note that we can ignore the exponential damping part for $\vec L_{(l)}$.

For $m>1$ and $ \lambda_{\text{cross}} >> 1$, the angular integrals along the Fermi surface directions give a factor proportional to
\bqa
\label{ang1}
 \int_0^{\pi} d \theta \,
\frac {  {\bar {\cal D }} (\vec L, \vec P, \vec Q) \, \sin^{m-2} \theta}
{  {\bar {\cal D}}^2(\vec L, \vec P, \vec Q) + 4 (  |\vec L_{(l)}| \, |\vec L_{( q )}| \, \cos \theta )^2   } 
& \simeq &
 \frac{ \pi }  { 2 \, |\vec L_{(l)}|  \, |\vec L_{( q )}| } 
\eqa
in the limit $ \frac{ {\bar {\cal D }} (\vec L, \vec P, \vec Q)  }  { 2 \, |\vec L_{(l)}|  \, |\vec L_{( q )}| } << 1 $, which is valid when $ |\vec L_{( q )}|^2 >> \frac{ \Lambda } 
{ \left ( \lambda_{\text{cross}} \right )^{\frac{1} { m+1} }  }  $.
This follows from the fact that
the main contribution to the integral over $|\vec L_{(l)}|$ comes from 
$|\vec L_{(l)}| \sim \tilde \alpha^{\frac{ 1} { 3 } } \, |\vec L-\vec P|^{\frac{d-m}{3}}
>> \Lambda$ 
in the large $\lambda_{cross}$ limit.
Using
\beq
\label{llint}
\int   d |\vec L_{(l)}| \, \frac{|\vec L_{(l)}|^{m-1} } {|\vec L_{(l)}|^3  + \tilde \alpha \, |\vec L -\vec P|^{d-m}}
= \frac{\pi }  { 3\, \sin \left( \frac{m\,\pi} {3} \right) |\vec L - \vec P|^{\frac{(d-m)\, (3-m)}{3}}   \, \tilde \alpha^{\frac{ (3-m)}{3}}  } \quad \mbox{for} \, \, 0<m<3\,,
\eeq
we perform the $|\vec L_{(l)}|$-integral to obtain
\bqa
\Pi_2 (q) &\sim & 
-\frac{e^4 \mu^{2\, x} \, \pi}  { 48  \,  N}  \,\left(\frac{k_F}{12 \, \pi} \right)^{\frac {m-1} {2}}
\int  \frac{ d\vec L\,d\vec P}{(2\pi)^{ 2d- m }} \,
\frac{{\cal B}_{4}(\vec L, \vec P, \vec Q) }
{ {\cal D}_{4} (\vec L, \vec P, \vec Q) } \,
\frac{ \pi}
{ | \vec L_{(q)} |^2 \,  \tilde \alpha^{\frac{ 3-m }{3}}  \,|\vec L - \vec P|^{\frac{(d-m)\, (3-m)}{3} }
\sin\left( \frac{m\, \pi} {3} \right) } 
 \,.
\eqa
The total power of $e$ comes out to be $\frac{2 \, (m+3) }{3}$, and we find 
\beq
\Pi_2 (q)   \sim 
- \frac { \tilde{e}^{\frac{m} {m+1}} } {  k_F^{\frac{m-1}{2\,(m+1)} } }
\frac{\left( e^2 \, k_F^{\frac{m-1}{2} } \right) \,\pi^2} 
{ 48 \,N \,| \vec L_{(q)} |^2 \, \sin\left( \frac{m\, \pi} {3} \right) }
\int  \frac{ d\vec L\,d\vec P}{(2\pi)^{ 2d- m }} \,
\frac{ {\cal B}_{4}(\vec L, \vec P, \vec Q) }
{{\cal D}_{4} (\vec L, \vec P, \vec Q) \,
|\vec L - \vec P|^{\frac{(d-m)\, (3-m)}{3} }
 }
\sim 
\frac { \tilde{e}^{\frac{m} {m+1}} } {  k_F^{\frac{m-1}{2\,(m+1)} } }
\frac{ | \vec Q |^{\frac{  m } {m+1} } 
}
{ N \, | \vec L_{(q)} |  }   \, \Pi_1 (q)  \,,
\eeq
to leading order in $ |\vec Q | $ and $\epsilon$. 
Therefore, the two-loop diagram is suppressed 
not only by a higher power of $\tilde e$ but also by
$\left( \frac { |{\bf Q}| } {  k_F} \right)^{\frac{m-1}{2\,(m+1)} }$
compared to the one-loop diagram.


\subsection{Two-loop contribution to fermion self-energy}
\label{2loopferm}

The two-loop fermion self-energy 
in Fig.~\ref{fig:ferm2}(a) 
is given by
\beq\label{twoloopf}
\Sigma_2 (q) = \frac{(ie)^4 \mu^{2\,x}}{N^2} 
\int {dp \,dl} \, D_1 (p)\, D_1 (l) \,
\gamma_{d-m} \, G_0 (p+q) \, \gamma_{d-m} \, G_0 (p+l+q) \, \gamma_{d-m} \, G_0 (l+q) \, \gamma_{d-m} \,.
\eeq
Using the gamma matrix algebra, we find that 
the self-energy can be divided into two parts:
\beq
\Sigma_2 (q) = \Sigma_{2a} (q) + \Sigma_{2b} (q) \,,
\eeq
where
\bqa
\Sigma_{2a,2b} (q) &=& \frac{i \, e^4 \, \mu^{2\,x}}{N^2} 
\int {dp \,dl} \,
 D_1 (p) \,D_1 (l) \frac{{\cal C}_{a,b}}{[(\vec P +\vec Q)^2 +\delta_{p+q}^2]\,
[(\vec P +\vec L+ \vec Q)^2 +\delta_{p+l+q}^2]\,
[(\vec L +\vec Q)^2 +\delta_{l+q}^2] } \,, \nn
\eqa
with
\bqa
{\cal C}_{a} & = &\gamma_{d-m} \Big [ \delta_{p+q} \,\delta_{p+l+q} \, \delta_{l+q} -
\delta_{l+q} \, \lbrace \, \vec \Gamma \cdot (\vec P +\vec Q)\rbrace
\, \lbrace \,  \vec \Gamma \cdot (\vec P +\vec L+ \vec Q) \rbrace 
-  \delta_{p+q} \, \lbrace \,  \vec  \Gamma \cdot (\vec P +\vec L+ \vec Q)  \rbrace
\, \lbrace \,  \vec \Gamma \cdot (\vec L +\vec Q) \rbrace  \nn
&&\qquad  \, \, \, \, - \delta_{p+l+q} 
\, \lbrace \,  \vec \Gamma \cdot (\vec P +\vec Q) \rbrace
\, \lbrace \,  \vec \Gamma \cdot (\vec L +\vec Q) \rbrace  \Big ] \,, \\
{\cal C}_{b} & = & [ \, \vec \Gamma \cdot (\vec P +\vec Q) \, ] \,
[ \, \vec \Gamma \cdot (\vec P +\vec L+ \vec Q) \, ] \,
[ \, \vec \Gamma \cdot (\vec L +\vec Q) \, ] \
- \delta_{p+q} \,\delta_{l+q}\,  [\vec \Gamma \cdot (\vec P +\vec L+ \vec Q)] - \delta_{p+l+q} \,\delta_{l+q} 
\, [ \, \vec \Gamma \cdot (\vec P +\vec Q)\, ]\nn
&& \,   \,  - \delta_{p+q} \,\delta_{p+l+q} \, [ \, \vec \Gamma \cdot (\vec L +\vec Q) \, ] \,.
\eqa
Shifting the variables as
\beq
p_{d-m} \rightarrow  p_{d-m} -\delta_q -2\,  \vec L_{(p)} \cdot \vec L_{(q)} -  \vec L_{(p)}^2
\,, \quad
l_{d-m} \rightarrow  l_{d-m} -\delta_q -2 \, \vec L_{(l)} \cdot \vec L_{(q)} -  \vec L_{(l)}^2 \,, \nonumber
\eeq
and integrating over $p_{d-m}$ and $l_{d-m}$, we obtain
\bqa
\Sigma_{2a} (q) & = & \frac{ i \, e^4 \, \mu^{2\,x}}  {4 \,N^2} 
\int \frac{d\vec P d\vec L}{(2\pi)^{2d-2m}} \frac{d \vec L_{(p)} \,  d \vec L_{(l)}}{(2\pi)^{2m}} \,
\frac{\gamma_{d-m} \,(\delta_q -2\,  \vec L_{(l)} \cdot \vec L_{( p )} )  \, 
{\bar {\cal C}}_a (\vec L, \vec P, \vec Q)   \,D_1 (p)\, D_1 (l)  }
{ (\delta_q -2\, \vec L_{(l)} \cdot \vec L_{(p)} )^2 + {\bar {\cal C}} (\vec L, \vec P, \vec Q) ^2   } \,,  \label{2loosiga} \\
\Sigma_{2b} (q) & = & \frac{ i \, e^4 \, \mu^{2\,x}}  {4 \,N^2}  
\int \frac{d\vec P d\vec L}{(2\pi)^{2d-2m}} \frac{d \vec L_{(p)} \,  d \vec L_{(l)}}{(2\pi)^{2m}} \,
 \frac{  {\bar {\cal C}} (\vec L, \vec P, \vec Q) \,
{\bar {\cal C}}_b  (\vec L, \vec P, \vec Q)   \,D_1 (p) \,D_1 (l) }
{(\delta_q -2\, \vec L_{(l)} \cdot \vec L_{(p)} )^2 +  {\bar {\cal C}} (\vec L, \vec P, \vec Q) ^2   } \,, \label{2loosigb}
\eqa
where 
\bqa
{\bar {\cal C}} (\vec L, \vec P, \vec Q) &=& |\vec P +\vec Q| + |\vec L +\vec Q|
+|\vec P +\vec L +\vec Q | \,,\nn
{\bar {\cal C}}_a (\vec L, \vec P, \vec Q) & = &
1-\frac{ [\vec \Gamma \cdot (\vec P +\vec Q)] \,
[\vec \Gamma \cdot (\vec  P +\vec L +\vec Q)]}{|\vec P +\vec Q|\,
|\vec P +\vec L +\vec Q|}  - \frac{ [\vec \Gamma \cdot (\vec P +\vec L +\vec Q)] \,
[\vec \Gamma \cdot (\vec L +\vec Q)]}{|\vec P +\vec L + \vec Q|\,
|\vec L +\vec Q|} + 
\frac{ [\vec \Gamma \cdot (\vec P +\vec Q)] \,
[\vec \Gamma \cdot (\vec L +\vec Q)]}{|\vec P +\vec Q|\,
|\vec L +\vec Q|} \,, \nn
{\bar {\cal C}}_b (\vec L, \vec P, \vec Q) & = &
\frac{ [\vec \Gamma \cdot (\vec P +\vec Q)] \,
[\vec \Gamma \cdot (\vec  P +\vec L +\vec Q)] \,
[\vec \Gamma \cdot (\vec L +\vec Q)]}
{|\vec P +\vec Q|\, |\vec P +\vec L +\vec Q| \, |\vec L +\vec Q|} 
- \frac{[\vec \Gamma \cdot (\vec L +\vec Q)]}{|\vec L +\vec Q|}
+ \frac{[\vec \Gamma \cdot (\vec L + \vec P+ \vec Q)]}
{|\vec L +\vec P+ \vec Q|}
- \frac{[\vec \Gamma \cdot (\vec P +\vec Q)]}{|\vec P +\vec Q|} \,.
\label{cab}\nn
\eqa
We can extract the UV divergent parts by expanding the integrand for small $\delta_q$ in Eq.~(\ref{2loosiga}), and  by setting $\delta_q =0$ in Eq.~(\ref{2loosigb}).  In this limit, we have:
\bqa
\Sigma_{2a} (q) & = & \frac{ i \, e^4 \, \mu^{2\,x} }  {4 \,N^2} 
\int \frac{d\vec P \, d\vec L\, d \vec L_{(p)} \,  d \vec L_{(l)} }{(2\pi)^{2d}} 
\frac{  \gamma_{d-m} \, \delta_q \,{\bar {\cal C}}_a (\vec L, \vec P, \vec Q) \,D_1 (p)\, D_1 (l) \,
\lbrace{\bar {\cal C}} (\vec L, \vec P, \vec Q) ^2 - 4\, ( \, \vec L_{(l)} \cdot \vec L_{(p)} )^2  \rbrace
 }
{ \lbrace  {\bar {\cal C}} (\vec L, \vec P, \vec Q) ^2 + 4\, ( \, \vec L_{(l)} \cdot \vec L_{(p)} )^2  \rbrace^2   } \,,   \\
\Sigma_{2b} (q) & = & \frac{ i \, e^4 \, \mu^{2\,x}}  {4 \,N^2}  
\int \frac{d\vec P d\vec L}{(2\pi)^{2d-2m}} \frac{d \vec L_{(p)} \,  d \vec L_{(l)}}{(2\pi)^{2m}} \,
 \frac{{\bar {\cal C}} (\vec L, \vec P, \vec Q) \, \,
{\bar {\cal C}}_b  (\vec L, \vec P, \vec Q)   \,D_1 (p) \,D_1 (l) }
{4 \,( \vec L_{(l)} \cdot \vec L_{(p)} )^2 +  {\bar {\cal C}} (\vec L, \vec P, \vec Q) ^2   } \,.
\eqa
For $m>1$, the angular integrals along the Fermi surface directions give a factor proportional to
\bqa
\label{angcase}
 \int_0^{\pi} d \theta \,
\frac {  {\bar {\cal C}}^2 (\vec L, \vec P, \vec Q)
- 4 \left ( \, |\vec L_{(l)}| \,| \vec L_{(p)}| \, \cos \theta \right )^2 }
{
\Big  \lbrace  {\bar {\cal C}} (\vec L, \vec P, \vec Q) ^2 
+ 4 \left ( \, |\vec L_{(l)}| \,| \vec L_{(p)}| \, \cos \theta \right )^2  
\Big \rbrace^2   }
\sin^{m-2} \theta 
 \simeq 
\begin{cases}
 \frac{ \sqrt{\pi}  \,  \Gamma \left( \frac{m-1}{2} \right )  }
 {  {\bar {\cal C}}  ^2 \, \Gamma \left( \frac{m}{2} \right ) } \,, 
& \text{for}\ 
\frac{ {\bar {\cal C}} (\vec L, \vec P, \vec Q)} {2\, |\vec L_{(l)}|  \, |\vec L_{(p)}| } >>1 \,,\nonumber \\
\frac{ \sqrt{\pi}  \,  \Gamma \left( \frac{m-1}{2} \right )  }
 {  2 \, \vec L_{(l)}^2 \, \vec L_{(p)}^2 \, \Gamma \left( \frac{m}{2} -1  \right ) }
+ \frac{\pi  \,(3-m) \,  {\bar {\cal C}}  } { 8 \, |\vec L_{(l)}|^3  \, |\vec L_{(p)}|^3 } 
 \,, 
& \text{for}\ 
\frac{ {\bar {\cal C}} (\vec L, \vec P, \vec Q)} {2\, |\vec L_{(l)}|  \, |\vec L_{(p)}| } << 1\,,
\end{cases} \nn
\eqa
for $\Sigma_{2a}$;
and
\bqa
 \int_0^{\pi}  d \theta \,
\frac {  {\bar {\cal C }} (\vec L, \vec P, \vec Q) \, \sin^{m-2} \theta}
{  {\bar {\cal C}}^2(\vec L, \vec P, \vec Q) 
+ 4 \left ( \, |\vec L_{(l)}| \,| \vec L_{(p)}| \, \cos \theta \right )^2   } 
& \simeq &
\begin{cases}
 \frac{ \sqrt{\pi}  \,  \Gamma \left( \frac{m-1}{2} \right )  }
 {  {\bar {\cal C}}  \, \Gamma \left( \frac{m}{2} \right ) } \,, 
& \text{for}\ 
\frac{ {\bar {\cal C}} (\vec L, \vec P, \vec Q)} {2\, |\vec L_{(l)}|  \, |\vec L_{(p)}| } >>1 \,,\nonumber \\
 \frac{ \pi }  { 2 \, |\vec L_{(l)}|  \, |\vec L_{(p)}| }
- \frac{ \sqrt{\pi}  \, {\bar {\cal C }}\, \Gamma \left( \frac{m-1}{2} \right )  }
 {  2 \, \vec L_{(l)}^2 \, \vec L_{(p)}^2 \, \Gamma \left( \frac{m}{2} -1  \right ) }
  \,, 
& \text{for}\ 
\frac{ {\bar {\cal C}} (\vec L, \vec P, \vec Q)} {2\, |\vec L_{(l)}|  \, |\vec L_{(p)}| } << 1 \,,
\end{cases} \nn
\eqa
for $\Sigma_{2b }$.
For $\frac{ {\bar {\cal C}} (\vec L, \vec P, \vec Q)} {2\, |\vec L_{(l)}|  \, |\vec L_{(p)}| } << 1$, 
the leading and the second leading order terms 
in $\frac{ {\bar {\cal C}} (\vec L, \vec P, \vec Q)} {2\, |\vec L_{(l)}|  \, |\vec L_{(p)}| }$ are kept.
Although the second leading term is not important for $\Sigma_{2 b} $,
it plays an important role for $\Sigma_{2 a}$ at  $m=2$
because the leading term vanishes at $m=2$.

Let us first estimate $\Sigma_{2a}$.
It is convenient to perform the integrations for 
$\vec L_{(l)}$ and 
$\vec L_{(p)}$ 
in the four regions separately, 
\bqa
\Bigg\{
0 < | \vec L_{( l )}| < 
\frac{ {\bar {\cal C}}  } 
{  2 \, |\vec L_{( p )} | } , &&
~~~~\, 0 < | \vec L_{( p )}| < 
\frac{ {\bar {\cal C}}  } 
{  2 \,\tilde \alpha^{ \frac{ 1 } {3} } \, |\vec L |^{  \frac{d-m}{3} } } \Bigg\}, \nn
\Bigg\{ 0 < | \vec L_{( l )}| < 
\frac{ {\bar {\cal C}} }   { 2\, |\vec L_{( p )}| } , &&
~~~~\, 
\frac{ {\bar {\cal C}}  } 
{  2 \,\tilde \alpha^{ \frac{ 1 } {3} } \, |\vec L |^{  \frac{d-m}{3} } }
<  | \vec  L_{( p )} | < \infty \Bigg\}, \nn
\Bigg\{ \frac{ {\bar {\cal C}} } 
{ 2\, |\vec L_{( p )}| } < | \vec L_{( l )}| 
< \infty , &&
~~~~\, 
0 < | \vec L_{( p )}|  < 
\frac{ {\bar {\cal C}}  } 
{  2 \,\tilde \alpha^{ \frac{ 1 } {3} } \, |\vec L |^{  \frac{d-m}{3} } } \Bigg\},  \nn
\Bigg\{ \frac{ {\bar {\cal C}} } 
{ 2\, |\vec L_{( p )}| } < | \vec L_{( l )}| 
< \infty , &&
~~~~\,  \frac{ {\bar {\cal C}}  } 
{  2 \,\tilde \alpha^{ \frac{ 1 } {3} } \, |\vec L |^{  \frac{d-m}{3} } }
< | \vec L_{( p )}| < \infty \Bigg\} . 
\eqa
Adding the result of integrations in each region, we obtain
\bqa
\Sigma_{2a} (q) & \sim & \frac{ i \, e^4 \, \mu^{2\,x}\, \gamma_{d-m} \, \delta_q \, \pi }  { 4 \, N^2} 
\int \frac{d\vec P d\vec L}{(2\pi)^{ 2d }}\, 
{\bar {\cal C}}_a (\vec L, \vec P, 0 ) 
\Big [\,
 \frac { \pi^{\frac{3} {2}} \, \Gamma \left(  \frac{m-1} {2} \right)
 \, \sec ^2 \left ( \frac{ ( 2 m+1)\, \pi } { 3 } \right) }
{ 18  \, \Gamma \left(  \frac{m} {2}  -1 \right) \,\tilde \alpha^{ \frac {  2\, ( 4-m) } {3} } \,( |\vec L |\,|\vec P| )^{ \frac { (d-m)\,( 4-m) } {3} }     } \nn
&&  \qquad \qquad \qquad \qquad    + \,
\frac{
{\bar {\cal C}}^{m-1} (\vec L, \vec P, 0 )   }
{ 8\, \tilde \alpha^{2} \, \left(\,|\vec L|\, |\vec P|\, \right)^{ d-m}   }
\Big \lbrace \,
\frac{  2^{2-m}
\left  ( \, m+4  - 3\,(m+1) \, \log \mathcal A \,  \right) 
\, \Gamma \left( \frac{m-1}{2}\right)  }   { 3\, \sqrt{\pi}  \, (m+1)^2\, \Gamma \left(\frac{m}{2} \right) }  
+ \, \frac { \frac{\mathcal A^{2-m} -1  }{2-m} -   \log \mathcal A } 
{ 2^{ m-2 } (2-m)/(3-m)} \nn
&&  \qquad \qquad \qquad \qquad
 + \, \frac{ 2^{2-m}  \, (3-m)  } {3} \,  
 \frac {  1  -   \frac{ 3\, \mathcal A^{2-m} }{5-m}  } 
{ 2-m}
 + \, \frac{3-m} {5-m} \, \frac {  2^{2-m}  \left ( 1-\mathcal A^{ 2-m } \right ) } 
{ 2-m } 
+ \frac { 2^{2-m} 
\mathcal (3-m) \,( 1 + 3\, \mathcal A ^{2-m} )  }  { 3\, (5-m)^2 }
\Big \rbrace 
\Big]\,,\nn
\eqa
where
\beq
\mathcal{A} (\vec L, \vec P, \alpha )
=\frac{\bar {\cal C} (\vec L, \vec P,0 ) } 
{ 2 \left(\,|\vec L|\, |\vec P|\, \right)^{\frac{d-m}{3} }  \tilde \alpha^{\frac{2}{3}} }.
\label{A}
\eeq
Here $ \vec Q$ dependent terms are dropped because they are sub-leading compared to the one that depends on $\delta_q $.
Note that $\frac {\mathcal A^{2-m} -1}  {2-m}$ and
$ \frac { \frac{\mathcal A^{2-m} -1  }{2-m} -   \log \mathcal A } { 2^{ m-2 } (2-m)/(3-m)}$ 
become terms that include $\log \mathcal A$
and  $\log^2 \mathcal A$ at $m=2$. 
From similar integrations, $\Sigma_{2b} (q)$ is evaluated to be
\bqa
\Sigma_{2b} (q) & \sim & \frac{ i \, e^4 \, \mu^{2\,x} \pi^3 }  {  72 \,N^2}  
\int \frac{d\vec P d\vec L}{(2\pi)^{ 2d }} \,
\frac{ {\bar {\cal C}}_b  (\vec L, \vec P, \vec Q)   }
{\tilde \alpha^{\frac{ 2\,(3-m)}{3}}  \left( |\vec L | \, |\vec P|\right )^{\frac{(d-m)\, (3-m)}{3} }
\sin^2 \left( \frac{m\, \pi} {3} \right) }
\,.
\eqa

For $d=d_c-\epsilon$, we find that
\bqa
&& \Sigma_{2a} (q)  \sim  \frac{ i  \, \gamma_{d-m} \, \delta_q\,\pi  }  { 4 \, N^2} 
\int \frac{d\vec P d\vec L}{(2\pi)^{ 2d }}\, 
{\bar {\cal C}}_a (\vec L, \vec P, 0) 
\, \Big [\,\frac {{\tilde e}^{\frac{ 2\,(m-1) } {m+1}} } { k_F^{\frac{ 2\,(m-1)} { m+1  } } }
\frac { \pi^{ \frac{3}{2} }\,\, \Gamma \left(  \frac{m-1} {2} \right) 
 \, \sec ^2 \left ( \frac{ ( 2 m+1)\, \pi } { 3 } \right)  }
{ 18 \, \Gamma \left(  \frac{m} {2} -1  \right)  \, ( |\vec L |\,|\vec P | )^{ \frac { (d-m)\,(  4-m) } {3} }    } 
\nn
&&
+
\frac{
  { \bar {\cal C}}^{m-1} (\vec L, \vec P, 0)   }
{ 8 \, k_F^{m-1}  \left(\,|\vec L|\, |\vec P|\, \right)^{ d-m}   } 
\Big \lbrace \,
\frac{  2^{2-m}  
\left  ( \, m+4  - 3\,(m+1) \, \log \mathcal A \,  \right)
\, \Gamma \left( \frac{m-1}{2}\right)  }   { 3\, \sqrt{\pi}  \, (m+1)^2\, \Gamma \left(\frac{m}{2} \right) } 
+ \, \frac { \frac{  
\left(
\frac{ \mathcal A |_{ \tilde \alpha =1 }} {{\tilde e}^{\frac{2 } {m+1} }   \, k_F^{  \frac{ m-1 } {m+1}  } 
\, \beta_d^{\frac{2}{3}} } 
\right)^{2-m} 
 -1  }{2-m} -   \log \mathcal A } 
{ 2^{ m-2 } (2-m)/(3-m) } \nn
&&  
+ \, \frac{ 2^{2-m} \, (3-m)  } {3}  
\,\frac {  1 -   \frac{ 3 } {5-m} 
  \left(
\frac{ \mathcal A |_{ \tilde \alpha =1 }} {{\tilde e}^{\frac{2 } {m+1} }   \, k_F^{  \frac{ m-1 } {m+1}  } 
\, \beta_d^{\frac{2}{3}}   } 
\right)^{2-m} 
  } 
{ 2-m} 
+ \, \frac{3-m} { 5-m }  \,
\frac {  1- 
  \left(
\frac
{ \mathcal A |_{ \tilde \alpha =1 }}
{{\tilde e}^{\frac{2 } {m+1} }   \, k_F^{  \frac{ m-1 } {m+1}  } 
\, \beta_d^{\frac{2}{3}}   } 
\right)^{ 2-m }   } 
{ 2^{ m-2 } \,(2-m) }
+ \frac {  1 + 3\, 
\left(
\frac
{ \mathcal A |_{ \tilde \alpha =1 }}
{{\tilde e}^{\frac{2 } {m+1} }   \, k_F^{  \frac{ m-1 } {m+1}  } 
\, \beta_d^{\frac{2}{3}}   } 
\right)^{ 2-m }    
}  
{2^{ m-2 } \times  3\, (5-m)^2 / (3-m)}
\Big \rbrace  \,
\Big]\,, \nn
\label{b39}
\eqa
and
\bqa
\Sigma_{2b} (q) & \sim & 
\frac {{\tilde e}^{\frac{ 2m } {m+1}} } { k_F^{\frac{m-1} { m+1  } } }
\frac{ i \, \pi^3 }  {  72\,  N^2}  
\int \frac{d\vec P d\vec L}{(2\pi)^{ 2d }}
\frac{ {\bar {\cal C}}_b  (\vec L, \vec P, \vec Q) }
{ \left( |\vec L | \,  |\vec P|\right )^{\frac{(d-m)\, ( 3 - m)} {3} }
\sin^2 \left( \frac{m\, \pi} {3} \right) }
\sim 
{\tilde e}^{\frac{ 2m } {m+1}}
\left ( 
\frac{  \Lambda }
{ k_F } \right ) ^{ \frac{ m-1 } {m+1} } 
\frac{  i\, \vec \Gamma \cdot \vec Q } 
{ N ^2 } \,,
\label{b40}
\eqa
to leading order in $ \delta_q$, $\vec Q $ and $\epsilon$.
The two-loop fermion self-energy is suppressed by $\tilde e$ and  $\frac{\Lambda}{k_F}$
at low energies for  $m>1$.
Due to the Ward identity,
the two-loop vertex corrections 
shown in Fig.~\ref{fig:vert2} are also suppressed. 
It is noted that Eqs. (\ref{b39}) and (\ref{b40}) are finite in the $\epsilon \rightarrow 0$ limit.
This is because the Fermi energy $k_F$ enters
as a dimensionful parameter that further suppresses  
the two-loop contributions compared to the one-loop diagrams. 
Therefore the critical exponents do not receive quantum corrections from the two-loop diagrams
in the low-energy limit.

\end{widetext}

\end{document}